\documentclass[superscriptaddress,amsmath,amssymb,prd,preprintnumbers,showpacs,twocolumn]{revtex4-1}
\usepackage{graphicx}
\usepackage[T1]{fontenc} 
\usepackage{amssymb,amsmath,bm,natbib}
\usepackage{color}
\usepackage{slashed}
\usepackage{graphics}
\usepackage{graphicx}
\usepackage[utf8]{inputenc}
\usepackage[caption=false]{subfig}
\usepackage{hyperref}
\usepackage{url}
\usepackage{dsfont}
\usepackage{float}
\usepackage{cancel}
\usepackage{units}
\usepackage{blindtext}
\usepackage[utf8]{inputenc}
\usepackage{upgreek}
\usepackage{booktabs}
\usepackage[dvipsnames,table,xcdraw]{xcolor}
\usepackage{enumerate}
\usepackage{mathtools}
\usepackage{soul,color}
\usepackage[normalem]{ulem}
\begin{document}
\title{The boundary of the gravitational Standard-Model Extension}
\author{Carlos M. Reyes}
\email{creyes@ubiobio.cl}
\affiliation{Centro de Ciencias Exactas, Universidad del B\'{i}o-B\'{i}o, Casilla 447, Chill\'{a}n, Chile}
\author{Marco Schreck}
\email{marco.schreck@ufma.br}
\affiliation{Departamento de F\'{i}sica, Universidade Federal do Maranh\~{a}o, 
Campus Universit\'{a}rio do Bacanga, S\~{a}o Lu\'{i}s (MA), 65085-580, Brazil}
\begin{abstract}
A modification of General Relativity that is based on the  gravitational Standard-Model Extension and incorporates
nondynamical background fields has recently been studied via the ADM formalism.
Our objective in this paper is to develop a better understanding of the additional contributions that arise on the 
spacetime boundary $\partial\mathcal{M}$. An extension of the previously introduced boundary terms, which are
relevant in the context of asymptotically flat spacetimes, follows from the decomposition of $\partial\mathcal{M}$
into timelike and spacelike hypersurfaces. Furthermore, we present an alternative method of deriving the
field equations satisfied by the induced metric on the purely spacelike hypersurfaces of the foliated spacetime.
This leads to the dynamical part of the Einstein equations modified by the background fields. Our results 
have the potential to be applicable in various contexts such as modified black holes and cosmology.
\end{abstract}
\maketitle
\section{Introduction}
\label{sec:introduction}
The abundance of experimental tests of General Relativity 
(GR)~\cite{Einstein:1915bz,Dyson:1920cwa,Pound:1960zz,Everitt:2011hp,Will:2014kxa,Ciufolini:2019ezb} 
carried out for more than 100 years has demonstrated that GR provides a 
description of gravitational phenomena that works astoundingly well. The geometrization of 
gravity that Einstein envisioned also has a certain undeniable aesthetics to it and, so 
far, it has been impossible to adopt this description to the other fundamental interactions of nature. 
Despite the vast experimental support as well as the beauty of Einstein's 
gravity theory, even if one has the viewpoint that gravity stays classical 
at all length scales, GR exhibits at least some unsatisfactory properties.

For example, many physicists would agree that 
the occurrence of spacetime singularities in some solutions of GR, 
e.g., black-hole spacetimes is an issue that cannot simply be ignored. 
Both indirect observations of black-hole mergers via gravitational-wave detection by 
LIGO~\cite{LIGOScientific:2016aoc,LIGOScientific:2018mvr,LIGOScientific:2020ibl,LIGOScientific:2021djp}
and the impressive photographs of black-hole accretion disks in M87 and the 
Milky Way made by the EHT~\cite{EventHorizonTelescope:2019dse,EventHorizonTelescope:2022xnr} 
undoubtedly demonstrate that black holes are not merely mathematical 
vacuum solutions of 
the Einstein equations, but part of our reality. Thus, the proper 
understanding and treatment of black-hole singularities is paramount.

Cosmology reveals another possible issue 
of GR. Cosmological time evolution is largely affected by the 
gravitational pull of the matter content of our Universe. 
Therefore, GR forms the theoretical foundation of the current 
cosmological standard model, $\Lambda$CDM. Measurements of the large-scale structure of our
Universe \cite{SDSS:2005xqv,Daniel:2008et,BOSS:2016wmc}
and the precise mapping of the cosmological microwave background 
radiation \cite{WMAP:2003elm,WMAP:2008lyn,WMAP:2010qai,WMAP:2012nax,Planck:2015fie,Planck:2018vyg}
hint towards the existence of a completely mysterious 
entity known as Dark Energy~\cite{Turner:1998mg}, 
which is needed to account for the accelerated expansion of the Universe. 
Nothing whatsoever is known about the nature of Dark Energy and its 
physical properties, e.g., its negative pressure contradict the 
characteristics of any form of matter or energy that can be 
investigated in the laboratory. So it is needless to say that its 
introduction into cosmology is unsatisfactory. However, it could 
be the case that Dark Energy is only needed to make contact with 
measurements, since GR suffers severe alterations at the very large 
length scales that dominate cosmological late-time evolution.

These and other arguments suggest a refinement of GR, let it be at 
microscopic and/or cosmological scales. 
While a large number of modified-gravity theories has been proposed in
the literature~\cite{Heisenberg:2018vsk,Tasson:2016xib,Petrov:2020,Shankaranarayanan:2022wbx}, 
which are more or less well motivated, our article will be dedicated 
to a specific class of such theories. Our intention is to respect coordinate invariance as well as the 
full nonlinear structure of GR. Moreover, 
we will be working in a classical setting, i.e., no attempt is made 
to quantize gravity. Although extensions such 
as Finsler geometry~\cite{Finsler:1918,Antonelli:1993,Bao:2000}
could be considered, in principle, Riemannian geometry is maintained 
as the underlying geometrical foundation.

Instead, we give up one 
of the defining characteristics of Einstein's gravity, which is 
diffeomorphism invariance. The violation of the latter is 
parameterized by particular nondynamical background fields 
that are contained in the gravitational sector of the Standard-Model 
Extension (SME)~\cite{Kostelecky:2003fs,Bailey:2006fd,Bailey:2009me,Tso:2011up,Bailey:2013oda,Bailey:2014bta,Bonder:2015maa,Kostelecky:2016uex,Kostelecky:2020hbb,Kostelecky:2021tdf,Ivanov:2021bvk,Ye:2022yxr,Bonder:2021gjo}. 
This effective approach is comprehensive and parameterizes 
violations of diffeomorphism symmetry and local Lorentz 
invariance in gravity in a model-independent way. The SME is 
understood as a field theory framework that enables broad 
experimental tests of nonstandard gravitational physics such as diffeomorphism violation.
The yearly updated data tables~\cite{Kostelecky:2008ts} provide an extensive compilation of experimental constraints
on symmetry violation in gravity --- amongst the even larger set of bounds on Lorentz violation
in a nongravitational setting.

Recently, the Hamiltonian formulation~\cite{Arnowitt:1962hi,Misner:1973,Hanson:1976,Henneaux:1992,Carlip:1998,Bertschinger:2002,Thiemann:2007zz} has been 
developed for extensions~\cite{ONeal-Ault:2020ebv,Reyes:2021cpx,Reyes:2022mvm} of GR that 
exhibit diffeomorphism invariance breaking. Analyses of this kind rest upon the $(3+1)$ decomposition, which is often also
referred to as the ADM decomposition (formulation) according to the names of the physicists \cite{Arnowitt:1962hi} that
introduced this technique into GR. The latter is a formidable 
theoretical toolset being the base of advanced black-hole physics~\cite{Poisson:2002,Poisson:2004} as well as of numerical relativity~\cite{Gourgoulhon:2007ue,Font:2008fka,Baumgarte:2010,Gourgoulhon:2012}. It is one of the cornerstones of
powerful computer codes such as the \textit{Einstein Toolkit} \cite{Loffler:2011ay} and \textit{GRHydro} \cite{Mosta:2013gwu}
that solve highly complicated problems in GR numerically.

The ADM formulation has also proven 
to be a valuable technique to analyze modified-gravity theories from a formal perspective. For this reason 
it forms the technical foundation of the papers~\cite{ONeal-Ault:2020ebv,Reyes:2021cpx,Reyes:2022mvm}. 
In our current work, emphasis will be put on the behavior of the theory on the spacetime boundary. 
We intend to avoid integrations by parts, as these may imply essential contributions 
on the spacetime boundary that cannot simply be discarded. Furthermore, we will carry out 
a proper treatment of boundary terms that are of relevance in such an analysis.

One of the principal motivations to implement the ADM formulation in the context of the SME was to explore diffeomorphism 
violation in a strong-gravity regime complementing the studies within linearized 
modified gravity \cite{Kostelecky:2015dpa,Kostelecky:2016kfm,Seifert:2016tog,Kostelecky:2017zob,Seifert:2018mlk,Tasson:2018fzt,Mewes:2019dhj,Shao:2020shv,Nascimento:2021rlg,Wang:2021ctl,Wang:2020pgu,Zhao:2022pun,ONeal-Ault:2021uwu,Niu:2022yhr,Haegel:2022ymk}, in particular, on gravitational-wave
physics. The ADM formulation has also been fruitful to stimulate 
a new branch of research, which could be coined SME 
cosmology~\cite{Bonder:2017dpb,ONeal-Ault:2020ebv,Nilsson:2022mzq,Reyes:2022dil}. Moreover, 
this formalism enables the definition of a slew of important physical quantities 
such as the ADM mass \cite{Arnowitt:1961zz} or the ADM momentum \cite{Gourgoulhon:2007ue,Gourgoulhon:2012},
which are useful in, e.g., black-hole 
physics. So having the ADM-decomposed gravitational SME at someone's 
disposal, brings them into a position to study modified black holes. Finally, 
the canonical formulation of SME 
gravity could shed light on the possible 
issues related to the Bianchi identity of pseudo-Riemannian geometry 
in the context of explicit symmetry violation in gravity~\cite{Kostelecky:2003fs,Bluhm:2014oua,Bluhm:2016dzm,Bonder:2018asb,Bluhm:2019ato,Bonder:2020fpn,Kostelecky:2020hbb,Kostelecky:2021tdf,Bluhm:2021lzf} such as in Ho\v{r}ava-Lifshitz gravity \cite{Horava:2008ih,Horava:2009uw} (see also Refs.~\cite{Nilsson:2018knn,Nilsson:2019bxv,ONeal-Ault:2020ebv,Reyes:2022mvm}) and dRGT massive gravity~\cite{deRham:2010kj,deRham:2011rn,deRham:2014zqa,Kostelecky:2021xhb}.

The modified-gravity theory under consideration in Ref.~\cite{Reyes:2021cpx} was shown to require an extended 
Gibbons-Hawking-York (GHY) boundary term~\cite{York:1972sj,Gibbons:1976ue} involving the nondynamical background
fields. The introduction of such boundary terms~\cite{Reyes:2021cpx} prevents higher-order time derivatives of
the metric from occurring and, thus, they are crucial to ensure a well-defined principle of stationary action.
By doing so, the Hamiltonian of the modified-gravity theory was constructed and shown to be 
equivalent to the modified Einstein equations in the 
covariant approach \cite{Bailey:2006fd}, when these are projected onto spacelike hypersurfaces 
$\Sigma_t$ of the spacetime foliation~\cite{Reyes:2022mvm}.

The modern research program on spacetime boundaries in gravity was established by
the pioneering works of Arnowitt, Deser, and Misner~\cite{Arnowitt:1962hi} as well as Choquet-Bruhat~\cite{IDP}.
These papers laid the foundations for research on noncompact and asymptotically flat spacetimes $\mathcal{M}$,
which play a significant role, in particular, in the study of stars and black holes. 
Furthermore, the works of Gibbons, Hawking, and York~\cite{York:1972sj,Gibbons:1976ue} demonstrated the importance
and peculiarities of the variational formulation in gravity, which established a powerful approach for analyzing 
the physics on spacetime boundaries.

Other contexts that provide motivation for understanding boundary terms in gravity 
include the dynamics of binary systems and the gravitational 
waves they emit~\cite{Christodoulou:1987vv,LIGOScientific:2016aoc,Cutler:1994ys}, 
open inflation~\cite{Turok:1998he}, and the search for a theory of quantum 
gravity~\cite{Loop1,Loop2}. Moreover, in the setting of the AdS/CFT correspondence 
it is worthwhile to mention the regularization of the action in AdS 
spacetimes~\cite{ADSCFT,Witten}, extended regularization methods~\cite{Anastasiou:2020zwc} 
for the physical notion of mass and angular momentum~\cite{Poisson:2002,Poisson:2004}, 
black-hole physics~\cite{Hawking:1995fd}, formal derivation of the 
ADM energy in the limit
of asymptotically flat spacetimes~\cite{Chrusciel:1986xts}    
and extensions to non-orthogonal 
boundaries~\cite{Hawking:1996ww}. In general, a definition of physically meaningful conserved charges
in (asymptotically flat) spacetimes requires an averaging process over spatial and temporal
regions at infinity \cite{Arnowitt:1961zz,Brown:1986nw,Barnich:2001jy}. Hence, these quantities involve
surface integrals demonstrating how the properties of the gravitational system on spacetime
boundaries contain essential information.

In the current paper, we
focus on a specific form of the spacetime boundary $\partial\mathcal{M}$, which allows us to derive the 
dynamical field equations and to acquire an even better knowledge of the true role of the extended boundary terms. 
We will be obtaining a new set of boundary terms depending on the extrinsic curvature $k$ of two-dimensional 
hypersurfaces that give rise to a foliation of the timelike part of $\partial\mathcal{M}$.
The results are applicable in the context of black-hole physics modified by the presence of SME
background fields. A substantial amount of 
research~\cite{Casana:2017jkc,Colladay:2019lig,Ding:2019mal,Maluf:2020kgf,Gullu:2020qzu,Carvalho:2021jlp} 
has already been performed in this subarea,
which highlights that our approach and findings have the potential to be taken up by researchers in the future.

The paper is organized as follows. In Sec.~\ref{sec:boundary} we introduce 
the modified-gravity theory focused on, recapitulate some of its properties and 
define the notation to be used throughout the remainder of the article. Here, 
we also analyze the additional contributions on the spacetime boundary that emerge 
due to the presence of the SME background fields. Section~\ref{sec:modified-dynamics} 
is dedicated to deriving the dynamical field equations based on the findings in Sec.~\ref{sec:boundary}. 
A nontrivial shift vector will be included, which generalizes previous results. Finally, 
our findings will be concluded on in 
Sec.~\ref{sec:FinalRemarks}. Our metric signature is $(-,+,+,+)$ and we will employ natural 
coordinates with $c=1$ unless otherwise stated. As in our previous articles~\cite{Reyes:2021cpx,Reyes:2022mvm}, 
the \textit{Mathematica} package \textit{xTensor} \cite{xTensor:2020} provides significant computational support.
\section{The extended action}
\label{sec:boundary}
Consider the following modified Einstein-Hilbert (EH) action that involves 
a subset of coefficients of the minimal gravitational SME~\cite{Kostelecky:2003fs,Kostelecky:2020hbb}:
\begin{subequations}
\label{eq:modified-EH}
\begin{equation}
S_G=S_b+S_{\substack{\mathrm{ext} \\ \mathrm{GHY}}}   \,,
\end{equation}
with the bulk action
\begin{equation}
\label{eq:action-bulk}
S_b=\int_{\mathcal{M}}\mathrm{d}^4x\frac{\sqrt{-g}}{2\kappa}
\left[(1-u){}^{(4)}R+s^{\mu\nu}{}^{(4)}R_{\mu\nu}\right]\,,
\end{equation}
and the boundary action
\begin{equation}\label{Baction}
S_{\substack{\mathrm{ext} \\ \mathrm{GHY}}}=\oint_{\partial\mathcal{M}}   \mathrm{d}^3y   \,
\frac{\varepsilon \sqrt{q} } {2\kappa} \,    \left[2(1-u)K-s^{\mathbf{nn}}K+  K_{ab}s^{ab}\right]\,,
\end{equation}
\end{subequations}
with $\kappa=8\pi G_N$. We cover the four-dimensional spacetime manifold 
$\mathcal{M}$ with coordinates $x^{\mu}$ 
carrying Greek indices. As customary, $g_{\mu\nu}$ is the spacetime metric 
and $g:=\det(g_{\mu\nu})$ its determinant. 
Furthermore, ${}^{(4)}R_{\mu\nu}$ denotes the Ricci tensor and ${}^{(4)}R:={}^{(4)}R^{\mu}_{\phantom{\mu}\mu}$ the 
Ricci scalar on $\mathcal{M}$. The EH action is modified by a scalar 
background field $u$ and a tensor-valued one, which is called $s^{\mu\nu}$.  
The latter are nondynamical and lead to a breakdown of diffeomorphism 
invariance~\cite{Reyes:2021cpx,Reyes:2022mvm}.

To render Hamilton's principle well-defined, we also included an extended GHY boundary term~\cite{Reyes:2021cpx} where $q:=\det(q_{ab})$ 
is the determinant of the induced metric $q_{ab}$ on the boundary $\partial\mathcal{M}$ of $\mathcal{M}$. Generic coordinates
and indices are employed in Eq.~\eqref{Baction}, which will be made more explicit after decomposing $\partial\mathcal{M}$ into substantially
different parts below. The GHY 
action involves the extrinsic-curvature tensor $K_{ab}$ and the trace of the latter, $K:=K^a_{\phantom{a}a}=q^{ab}K_{ab}$. 
Moreover, $\varepsilon=\mp 1$ depending on whether $\partial\mathcal{M}$ is spacelike (timelike). Lightlike regions
on $\partial\mathcal{M}$ are sets of measure zero, which do not contribute to the surface integral in Eq.~\eqref{Baction}.
Moreover, setting all SME coefficients to zero, Eq.~\eqref{eq:modified-EH} reproduces the EH action with the GHY boundary
term, as expected.

\begin{figure}
\centering
\includegraphics[scale=0.65]{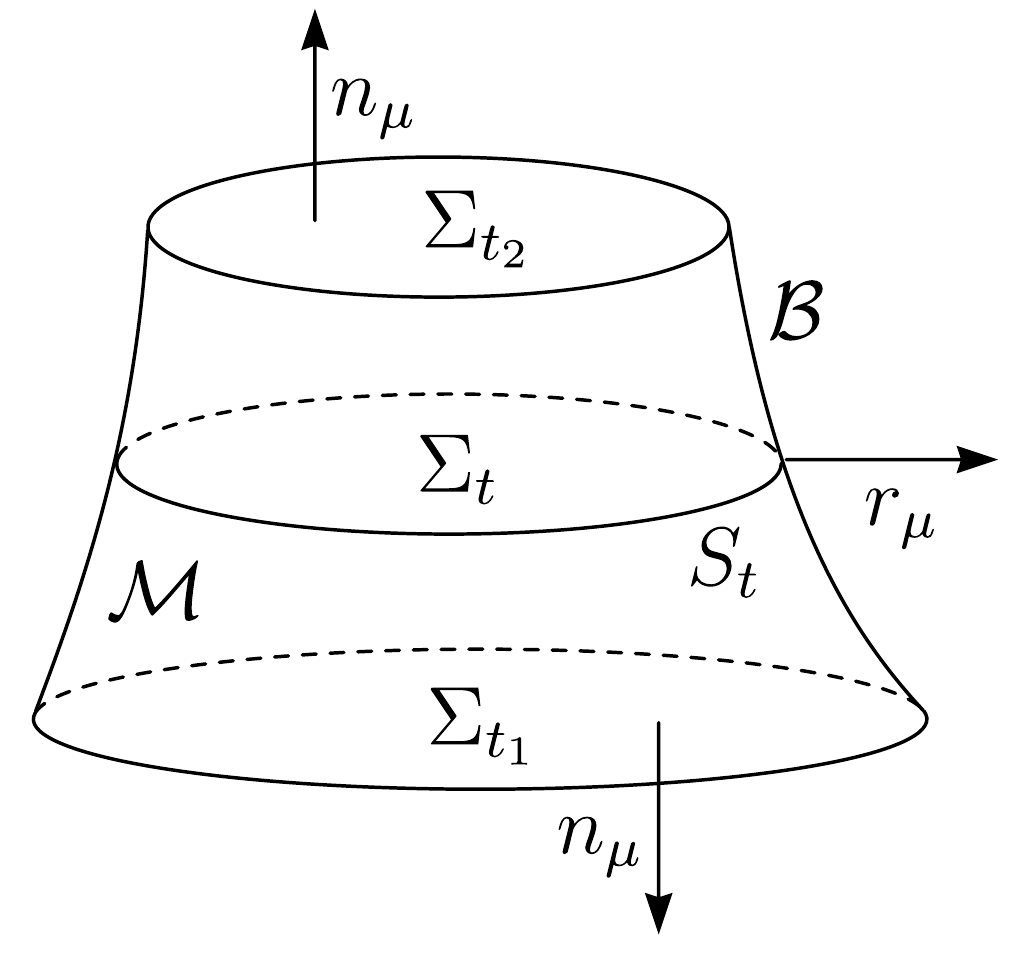}
\caption{Foliation of four-dimensional spacetime $\mathcal{M}$ in terms of embedded three-dimensional spacelike 
hypersurfaces $\Sigma_t$. The caps are formed by $\Sigma_{t_1}$ and $\Sigma_{t_2}$, respectively. The mantle 
$\mathcal{B}$ is foliated in terms of two-dimensional hypersurfaces $S_t$. Also, $n_{\mu}$ 
is normal to the caps and $r_{\mu}$ is orthogonal to the mantle.}
\label{fig:spacetime-boundary}
\end{figure}

Our first objective is to derive an ADM-decomposed action from Eq.~\eqref{eq:modified-EH}, which will be given by
Eq.~\eqref{eq:action-G-complete} towards the end of the current section. The machinery and procedure employed to arrive at
the latter are to be developed as follows. First of all, we focus on a spacetime $\mathcal{M}$ whose boundary $\partial\mathcal{M}$ 
is topologically a 3-cylinder, $\mathbb{R}\times S^2$, see Fig.~\ref{fig:spacetime-boundary}.
Let us foliate $\mathcal{M}$
in terms of 
spacelike hypersurfaces $\Sigma_t$ such that the boundary is expressed by $\partial\mathcal{M}=\Sigma_{t_1}
\cup \,\Sigma_{t_2}\cup\,\mathcal{B}$ with purely spacelike caps $\Sigma_{t_1},\Sigma_{t_2}$ and
a timelike mantle $\mathcal{B}$, according to Fig.~\ref{fig:spacetime-boundary}. 
For $\Sigma_t \subset \mathcal{M}$, which also includes $\Sigma_{t_1}$ and $\Sigma_{t_2}$, we 
consider coordinates $y^a$ with Latin indices $a,b,c,\dots$.

The foliation leads  to a natural decomposition of the 
tensor-valued background field $s^{\mu\nu}$ into three independent components:
\begin{equation}
\label{eq:decomposition-s}
s^{\alpha\beta}=q^{\alpha}_{\phantom{\alpha}\mu}q^{\beta}_{\phantom{\beta}\nu}s^{\mu\nu}
-(q^{\alpha}_{\phantom{\alpha}\nu}n^{\beta}+q^{\beta}_{
\phantom{\beta}\nu}n^{\alpha})s^{\nu \mathbf{n}}+n^{\alpha}n^{\beta}s^{\mathbf{nn}}\,,
\end{equation}
where $q^{\mu}_{\phantom{\mu}\nu}=\delta^{\mu}_{\phantom{\mu}\nu}+n^{\mu}n_{\nu}$ 
projects a part of a spacetime tensor described by a single Lorentz index onto $\Sigma_t$
and $n_{\mu}$ is a unit normal vector orthogonal to $\Sigma_t$.
To define the components of this decomposition, convenient for us, we introduce the valued 
tensors $e^{\mu}_a$, which are given by
\begin{align}
\label{eq:valued-tensors}
e^{\mu}_a:=\frac{\partial x^{\mu}}{\partial y^a} \,,
\end{align}
where the spacetime coordinates are understood to be parameterized as 
$x^{\mu}(y^a)$. Note that the $e^{\mu}_a$ govern pullback operations of 
covariant tensor fields~\cite{Gourgoulhon:2007ue,Gourgoulhon:2012,Blau:2020} 
that exist due to the embedding of $\Sigma_t$ into $\mathcal{M}$.
With this in mind, we define the tensor-valued purely spacelike part $s^{ab}$ of the background 
field through the relation
\begin{subequations}\label{eq:sectors-s}
\begin{equation}\label{eq:spacelike-s}
q^{\alpha}_{\phantom{\alpha}\mu}
q^{\beta}_{\phantom{\beta}\nu}s^{\mu\nu}=:e^{\alpha}_{a}e^{\beta}_{b}s^{ab}\,,
\end{equation}
and the scalar purely timelike contribution by
\begin{equation}\label{eq:timelike-s}
s^{\mathbf{nn}}:=s^{\mu\nu}n_{\mu}n_{\nu}\,.
\end{equation}
\end{subequations}
Since $s^{\mu\nu}$ and $s^{ab}$ are contravariant, by construction, Eq.~\eqref{eq:spacelike-s} 
cannot simply be solved for $s^{ab}$. Hence, $s^{ab}$ is defined implicitly by 
Eq.~\eqref{eq:spacelike-s} and the right-hand side of this relation can be interpreted as the pushforward
of $s^{ab}$ from $\Sigma_t$ into $\mathcal{M}$; see Eq.~(16.10) in Ref.~\cite{Blau:2020}. Then, 
$s^{ab}$ is understood as $s^{\mu\nu}$ suitably restricted to $\Sigma_t$
by the application of two valued tensors. It is also helpful to recall that 
$q^{\alpha \beta}=e^{\alpha}_ae^{\beta}_b  q^{ab} $, i.e., $q^{ab}$ can be lifted to $\mathcal{M}$
by a pushforward operation.

In principle, Eq.~\eqref{eq:decomposition-s} also contains a vector-valued mixed piece given 
by $s^{\mu\mathbf{n}}=:e^{\mu}_as^{a\nu}n_{\nu}$, but the latter can be gauged
away at first order in the coefficients~\cite{Reyes:2021cpx}, which is why we 
will discard $s^{\mu\mathbf{n}}$ in the following. Note also that in Ref.~\cite{Reyes:2021cpx}
we did not find any GHY-like boundary term associated with $s^{\mu\mathbf{n}}$, cf.~Eq.~\eqref{Baction}.

Moreover, it is reasonable to distinguish between quantities defined on spacelike 
and timelike hypersurfaces via different sets of indices. Therefore, 
let us introduce the following different submanifolds with their corresponding 
coordinates. For the mantle $\mathcal{B} \subset \partial\mathcal{M}$ we use coordinates $z^i$ 
and Latin indices $i,j,k,\dots$. For the closed two-surface 
$S_t \subset \Sigma_t$, which is the boundary of $\Sigma_t$,
we employ coordinates $\theta^A$ and capital Latin indices $A,B,C,\dots$.

The boundary action of Eq.~\eqref{Baction} is then decomposed~as
\begin{subequations}
\label{eq:decomposition-boundary-terms}
\begin{equation}
S_{\substack{\mathrm{ext} \\ \mathrm{GHY}}}=S_{\Sigma_{t_1}}+S_{\Sigma_{t_2}}+S_{\mathcal B}\,,
\end{equation}
with the contributions on the two caps and the mantle,
\begin{align}
\label{eq:action-sigma-1}
S_{\Sigma_{t_1}}&=\int_{\Sigma_{t_1}} 
\mathrm{d}^3y\,\frac{\sqrt{q}}{2\kappa}  \,\left[2(1-u)K-s^{\mathbf{nn}}K+K_{ab}s^{ab}\right]   \,, \\[2ex]
\label{eq:action-sigma-2}
S_{\Sigma_{t_2}}&=-\int_{\Sigma_{t_2}} 
\mathrm{d}^3y\,\frac{\sqrt{q}}{2\kappa} \,\left[2(1-u)K-s^{\mathbf{nn}}K+K_{ab}s^{ab}\right]\,, \\[2ex]
\label{eq:action-curly-B}
S_{\mathcal{B}}&=\int_{\mathcal{B}} \mathrm{d}^3z\, \frac{\sqrt{-\gamma}}{2\kappa} 
\,\left[2(1-u)\mathcal{K}+\mathcal{K}_{ij}s^{ij}\right]\,,
\end{align}
\end{subequations}
where the extrinsic curvature is defined appropriately on each hypersurface. 
In particular, on $\mathcal{B}$ we define the induced metric
$\gamma_{ij}:=g_{\alpha\beta}e^{\alpha}_{i}
e^{\beta}_{j}$ with $e^{\alpha}_{i}:=
\partial x^{\alpha}/\partial z^i$. We choose 
$r_a$ to be the unit normal to $S_t$ with associated four-vector $r^{\alpha}= r^a 
e_{a}^{\alpha}$ and the valued tensors $e^{\alpha}_{a}$ introduced previously in 
Eq.~\eqref{eq:valued-tensors}. Note that $r^{\alpha} n_{\alpha}=0$, as 
$r^{\alpha}$ is understood to live in $\Sigma_t$; cf.~Fig.~\ref{fig:spacetime-boundary}.
Furthermore, we define the extrinsic-curvature tensor on 
$\mathcal{B}$ as $\mathcal{K}_{ij}:=e^{\alpha}_{i}e^{\beta}_{j}\nabla_{\beta}r_{\alpha}$
where $\mathcal{K}:=\mathcal{K}^i_{\phantom{i}i}=\gamma_{ij}\mathcal{K}^{ij}$ is
its corresponding trace. The covariant derivative $\nabla_{\mu}$ 
is compatible with the metric $g_{\mu\nu}$ of $\mathcal{M}$.

Note also that, in principle, Eq.~\eqref{eq:action-curly-B} would contain a term 
proportional to $s^{\mathbf{rr}}\mathcal{K}$ with $s^{\mathbf{rr}}:=s^{\mu\nu}r_{\mu}r_{\nu}$. 
However, since $s^{\mathbf{nn}}$ provides a nonvanishing contribution for a timelike 
normal vector $n_{\mu}$ by its definition via Eq.~\eqref{eq:timelike-s}, it must hold that $s^{\mathbf{rr}}=0$ for a 
spacelike normal vector $r_{\mu}$ due to $r^{\mu}n_{\mu}=0$. The purely spacelike components 
of $s^{\mu\nu}$ are already contained in the term $\mathcal{K}_{ij}s^{ij}$ 
in Eq.~\eqref{eq:action-curly-B}, which is an implication of 
the way how $s^{\mu\nu}$ is decomposed in the foliation according to Eq.~\eqref{eq:decomposition-s}.

The key part of the forthcoming analysis is to focus on contributions providing total
derivatives compatible with $g_{\mu\nu}$ and $q_{ab}$, respectively. We will find that
the latter only occur for $s^{ab}$, which makes sense, as these coefficients result from
restricting $s^{\mu\nu}$ to the purely spacelike hypersurfaces $\Sigma_t$.
Let us now consider the decompositions (see, e.g., Refs.~\cite{Gourgoulhon:2007ue,Gourgoulhon:2012})
\begin{subequations}
\label{eq:decompositions-ricci}
\begin{align}
\label{eq:decomposition-ricci-scalar}
{}^{(4)}R&=R+K^2-K_{ab}K^{ab}-2R_{\mathbf{nn}}\,, \displaybreak[0]\\[2ex]
\label{eq:reformulation-Rnn}
R_{\mathbf{nn}}&=K^2-K_{ab}K^{ab}+\nabla_{\mu}\zeta^{\mu}\,, \displaybreak[0]\\[2ex]
\label{eq:decomposition-ricci-tensor}
q^{\beta}_{\phantom{\beta}\nu}q^{\delta}_{\phantom{\delta}\sigma}\prescript{(4)}{}
R_{\beta\delta}&=R_{\nu\sigma}+\nabla_{\mu}
\sigma^{\mu}_{\phantom{\mu}\nu\sigma}-a^{\beta}n_{\nu}K_{\beta\sigma} \nonumber \\
&\phantom{{}={}}-a^{\delta}n_{\sigma}K_{\nu \delta}-a_{\nu}a_{\sigma}-D_{\nu}a_{\sigma}\,,
\end{align}
\end{subequations}
where $R:=R^a_{\phantom{a}a}=q^{ab}R_{ab}$ is the Ricci scalar obtained from the trace of
the Ricci tensor $R_{ab}$ on $\Sigma_t$. Moreover, $D_{\mu}$ is the covariant derivative compatible with the intrinsic metric 
$q_{\mu\nu}$ and $a_{\alpha}:=n^{\mu}\nabla_{\mu}n_{\alpha}$ denotes the ADM acceleration.
In Eq.~\eqref{eq:decomposition-ricci-tensor} these quantities have been lifted to $\mathcal{M}$,
but a pullback onto $\Sigma_t$ can be performed via
$D_av_b=e_{a}^{\alpha} e_{b}^{\beta} D_{\alpha}v_{\beta}$ and $a_{c}=e_c^{\alpha}a_{\alpha}$
with the valued tensors of Eq.~\eqref{eq:valued-tensors}. We also defined the vector
\begin{subequations}
\label{eq:definitions-zeta-sigma}
\begin{equation}
\label{eq:definition-zeta}
\zeta^{\mu}:=n^{\nu}\nabla_{\nu}n^{\mu}-Kn^{\mu}\,,
\end{equation}
convenient to be used in Eq.~\eqref{eq:reformulation-Rnn}, as well as the third-rank tensor
\begin{equation}
\label{eq:definition-sigma}
\sigma^{\mu}_{\phantom{\mu}\nu\sigma}:=n^{\mu}K_{\nu\sigma}\,,
\end{equation}
which occurs in Eq.~\eqref{eq:decomposition-ricci-tensor}.
\end{subequations}
We emphasize that four-divergences of the latter quantities with
$g_{\mu\nu}$-compatible derivatives can be found in Eq.~\eqref{eq:decompositions-ricci}. 
These will play an important role below.

By applying the $(3+1)$ decomposition of the intrinsic curvature encoded in Eq.~\eqref{eq:decompositions-ricci} 
as well as the decomposition of $s^{\mu\nu}$ in Eq.~\eqref{eq:decomposition-s} --- with the mixed
coefficients omitted --- to Eq.~\eqref{eq:action-bulk} the bulk action can be ADM-decomposed as
\begin{align}
\label{eq:decomposition-Sb}
S_b&=\int_{\mathcal{M}} \mathrm{d}^4x\,\frac{\sqrt{-g}}{2\kappa}\bigg[(1-u)(R-K^2
+K_{ab}K^{ab}-2\nabla_{\mu}\zeta^{\mu}) \notag \\
&\phantom{{}={}}\hspace{1.8cm}+s^{\mathbf{nn}}n^{\mu}n^{\nu}{}^{(4)}R_{\mu\nu}
+s^{\mu\nu}q^{\beta}_{\phantom{\beta}\mu}q^{\delta}_{\phantom{\delta}\nu}{}^{(4)}R_{\beta\delta}\bigg]\,.
\end{align}
Hence, the ADM decomposition of the EH action is scaled by the factor $1-u$. Furthermore, 
the decomposition of $s^{\mu\nu}$ into purely timelike and spacelike 
parts, respectively, is evident. To rewrite the last two terms, we benefit 
from Eqs.~\eqref{eq:sectors-s}, \eqref{eq:reformulation-Rnn}, and \eqref{eq:decomposition-ricci-tensor} leading
to
\begin{align}\label{bbulk}
S_b&=\int_{\mathcal{M}} \mathrm{d}^4x\,\frac{\sqrt{-g}}{2\kappa}\Big[(1-u)
(R-K^2+K_{ab}K^{ab}-2\nabla_{\mu}\zeta^{\mu}) \notag \\[-1ex]
&\phantom{{}={}}\hspace{2.2cm}+s^{\mathbf{nn}}(K^2-K_{ab}K^{ab}+\nabla_{\mu}
\zeta^{\mu}) \notag \\
&\phantom{{}={}}\hspace{2.2cm}+s^{ab}(R_{ab}+e^{\nu}_a e^{\sigma}_b
\nabla_{\mu}\sigma^{\mu}_{\phantom{\mu}\nu\sigma} \nonumber \\
&\phantom{{}={}}\hspace{3.2cm}-a_{a}a_{b}-D_{a}a_{b})    \Big]\,,
\end{align}
which has now been expressed completely in terms of the components of $s^{\mu\nu}$ 
defined in Eqs.~\eqref{eq:spacelike-s}, \eqref{eq:timelike-s}. 
Then, the key terms giving rise to total derivatives 
in the bulk action~\eqref{bbulk}
are given by
\begin{align}
\label{eq:terms-providing-total-derivatives}
S_b & \supset 
 \int_{\mathcal{M}} \mathrm{d}^4x\,\frac{\sqrt{-g}}{2\kappa} \Big\{
\left[-2(1-u)+s^{\mathbf{nn}}\right]\nabla_{\mu}\zeta^{\mu} \notag \\
&\phantom{{}={}}\hspace{2.2cm}+s^{\nu\sigma}\nabla_{\mu}
\sigma^{\mu}_{\phantom{\mu}\nu\sigma} -  s^{ab}D_a a_b   \Big\}\,.
\end{align}
Note that the last term even contains a $q_{ab}$-compatible covariant derivative,
which is a property not to be encountered in the EH action.
Carrying out integrations by parts, the latter are written in the alternative form
\begin{align}
\label{eq:terms-providing-total-derivatives-reformulated}
S_b&\supset
\int_{\mathcal{M}} \mathrm{d}^4x\,\frac{\sqrt{-g}}{2\kappa}
\Big\{\nabla_{\mu}\left[-2(1-u)\zeta^{\mu}+s^{\mathbf{nn}}\zeta^{\mu}\right]\notag  \\
&\phantom{{}={}}\hspace{2.2cm}-\zeta^{\mu}\nabla_{\mu}(2u+s^{\mathbf{nn}}) \notag \\
&\phantom{{}={}}\hspace{2.2cm}+\nabla_{\mu}(s^{ab}\sigma^{\mu}_{\phantom{\mu}ab})
- \sigma^{\mu}_{\phantom{\mu}\nu \sigma} \nabla_{\mu}s^{\nu \sigma} \Big\} \notag \\
&\phantom{{}={}}+\int_{\mathcal{M}} \mathrm{d}t\mathrm{d}^3y\,\frac{\sqrt{q}}{2\kappa}
\left[-D_a (Ns^{ab}a_b)+ a_b D_a (Ns^{ab})\right]\,.
\end{align}
Here, we have used $s^{\nu \sigma}K_{\nu \sigma}=s^{ab}K_{ab}$, 
which can be proven from Eq.~\eqref{eq:spacelike-s}. 
The first and third lines of Eq.~\eqref{eq:terms-providing-total-derivatives-reformulated}
now involve $g_{\mu\nu}$-compatible 
total derivatives, whereas the last line contains a $q_{ab}$-compatible total 
derivative. These are complemented by correction terms such that 
Eq.~\eqref{eq:terms-providing-total-derivatives} can be reproduced neatly. 
Furthermore, the integral measure of the last line has been ADM-decomposed, since
the integrand only depends on properly ADM-decomposed quantities.

Gauss' theorem transforms the total derivatives in the first and third lines of
Eq.~\eqref{eq:terms-providing-total-derivatives-reformulated} into boundary terms:
\begin{align}
\label{eq:emergence-entire-boundary-term}
\int_{\mathcal{M}} &\mathrm{d}^4x\,\frac{\sqrt{-g}}{2\kappa}
\nabla_{\mu}\Big\{[-2(1-u)+s^{\mathbf{nn}}]\zeta^{\mu}+s^{ab}\sigma^{\mu}_{\phantom{\mu}ab}\Big\} \notag \\
&=\int_{\Sigma_{t_1}\cup \Sigma_{t_2}}\mathrm{d}^3y\,\frac{\sqrt{q}}{2\kappa}\,n_{\mu}\Big\{[-2(1-u)+s^{\mathbf{nn}}]\zeta^{\mu} \notag \\[-1ex]
&\phantom{{}={}}\hspace{4.85cm}+s^{ab}\sigma^{\mu}_{\phantom{\mu}ab}\Big\} \notag \\
&\phantom{{}={}}+\int_{\mathcal{B}}\mathrm{d}^3z\,\frac{\sqrt{-\gamma}}{2\kappa}\,r_{\mu}[-2(1-u)\zeta^{\mu}+s^{ab}\sigma^{\mu}_{\phantom{\mu}ab}] \notag \\
&=-\int_{\Sigma_{t_1}\cup \Sigma_{t_2}}\mathrm{d}^3y\,\frac{\sqrt{q}}{2\kappa}\Big\{[2(1-u)-s^{\mathbf{nn}}]K+s^{ab}K_{ab}\Big\} \notag \\
&\phantom{{}={}}+\int_{\mathcal{B}}\mathrm{d}^3z\,\frac{\sqrt{-\gamma}}{2\kappa}[-2(1-u)r_{\mu}n^{\nu}\nabla_{\nu}n^{\mu}]\,,
\end{align}
where we employed the definitions of Eq.~\eqref{eq:definitions-zeta-sigma}. We also benefited from the basic properties $n^2=-1$, $a_{\mu}n^{\mu}=0$, and $r_{\mu}n^{\mu}=0$ as well as $s^{\mathbf{rr}}=0$ on $\mathcal{B}$. Since Eq.~\eqref{eq:definition-zeta} depends on $a^{\mu}$ and $r_{\mu}a^{\mu}\neq 0$, the contribution on $\mathcal{B}$ stated in the last line of Eq.~\eqref{eq:emergence-entire-boundary-term} survives.

Now, the purely spacelike parts of Eq.~\eqref{eq:emergence-entire-boundary-term} given by the
two terms on the caps $\Sigma_{t_1}$  and $\Sigma_{t_2}$ cancel Eqs.~\eqref{eq:action-sigma-1}
and
\eqref{eq:action-sigma-2}, respectively, whereas the contributions on the mantle 
$\mathcal{B}$ will be treated later. As the term $D_a (Ns^{ab}a_b)$ involves a 
$q_{ab}$-compatible total derivative, it provides a boundary term on $\mathcal{B}$, 
which is already foliated in terms of two-dimensional hypersurfaces $S_t$ as follows:
\begin{align}
\int_{\mathcal{M}} \mathrm{d}t\mathrm{d}^3y\,&\frac{\sqrt{q}}{2\kappa}
[-D_a (Ns^{ab}a_b)] \notag \\
&=\int_{t_1}^{t_2} \mathrm{d}t \oint_{S_t}  
\mathrm{d}^2\theta \,\frac{N\sqrt{\sigma}}{2\kappa}(-r_is^{ij}a_j)\,.
\end{align}
Note that via a coordinate transformation, the indices $a,b$ are replaced by $i,j$ to represent coordinates 
on the foliated hypersurface $\mathcal B$. Here, $\sqrt{\sigma}$ is the integration measure
on $S_t$ that depends on the intrinsic metric $\sigma_{AB}$ on $S_t$. The latter
will be considered in more detail below.

The total action is then of the form
\begin{align}
\label{eq:total-action-decomposed-caps-canceled}
S_G&=\int_{\mathcal{M}} \mathrm{d}t\mathrm{d}^3y\,
\frac{N\sqrt{q}}{2\kappa}\bigg[(1-u)(R-K^2+K_{ab}K^{ab}) \notag \\
&\phantom{{}={}}\hspace{2.5cm}+s^{\mathbf{nn}}(K^2-K_{ab}K^{ab})+s^{ab}{R}_{ab} \notag \\
&\phantom{{}={}}\hspace{2.5cm}-s^{ab}a_{a}a_{b}-\zeta^{\mu}\nabla_{\mu}(2u+s^{\mathbf{nn}}) \notag \\
&\phantom{{}={}}\hspace{2.5cm}-\sigma^{\mu}_{\phantom{\mu}\nu\sigma}
\nabla_{\mu}s^{\nu\sigma}+\frac{1}{N}a_b D_a(Ns^{ab})\bigg] \notag \displaybreak[0]\\
&\phantom{{}={}}+\int_{\mathcal{B}} \mathrm{d}^3z\,\frac{\sqrt{-\gamma}}
{2\kappa}\Big[-2(1-u)r_{\mu}n^{\nu}\nabla_{\nu}n^{\mu}\Big] \notag \\
&\phantom{{}={}}-\int_{t_1}^{t_2}\mathrm{d}t\oint_{S_t}  
\mathrm{d}^2\theta \,\frac{N\sqrt{\sigma}}{2\kappa}r_ia_js^{ij} \notag \\
&\phantom{{}={}}+\int_{\mathcal{B}} \mathrm{d}^3z\, \frac{\sqrt{-\gamma}}{2\kappa} 
\,\left[2(1-u)\mathcal{K}+\mathcal{K}_{ij}s^{ij}\right]\,,
\end{align}
where the fifth and sixth lines contain the terms on the mantle $\mathcal{B}$ that remain 
after applying Gauss' theorem to each of the total derivatives in 
Eq.~\eqref{eq:terms-providing-total-derivatives-reformulated}. As mentioned before,
the contribution in the sixth line is already foliated properly in terms of $S_t$,
which will be helpful in the following. Note that the terms on $\mathcal{B}$ do not simply
cancel with the original boundary action $S_{\mathcal{B}}$ 
of Eq.~\eqref{eq:action-curly-B}, which has been reinstated explicitly into 
the seventh line of Eq.~\eqref{eq:total-action-decomposed-caps-canceled}.
A more sophisticated treatment of these contributions is indispensable, though.

Moreover, the bulk of Eq.~\eqref{eq:total-action-decomposed-caps-canceled} 
now depends on $g_{\mu\nu}$-compatible directional derivatives of the SME 
coefficients. It is beneficial to express these in terms of the ADM 
acceleration $a_c$ defined on $\Sigma_t$ and Lie derivatives~\cite{Carroll:1997ar} 
with respect to the vector $m^{\mu}:=Nn^{\mu}$:
\begin{subequations}
\begin{align}
\zeta^{\mu}\nabla_{\mu}u&=a^cD_cu-\frac{K}{N}\mathcal{L}_mu\,, \\[2ex]
\zeta^{\mu}\nabla_{\mu}s^{\mathbf{nn}}&=a^cD_cs^{\mathbf{nn}}-\frac{K}{N}\mathcal{L}_ms^{\mathbf{nn}}\,,
\\[2ex]
\label{eq:relation-sigma}
\sigma^{\mu}_{\phantom{\mu}\nu\sigma}\nabla_{\mu}s^{\nu\sigma}&=\frac 
1NK_{ab} \mathcal{L}_ms^{ab}+2K_{ab}K^a_{\phantom{a}c}s^{cb}\,.
\end{align}
\end{subequations}
The occurrence of Lie derivatives $\mathcal{L}_m$ of the background 
fields is characteristic when the ADM formalism is applied to sectors 
of the gravitational SME \cite{Reyes:2021cpx,Reyes:2022mvm}. For 
Eq.~\eqref{eq:relation-sigma} it is important to take into account 
that the Lie derivative along $m^{\mu}$
of a quantity living in $\Sigma_t$ remains in $\Sigma_t$.

The next step is to investigate the contributions on~$\mathcal{B}$. 
We intend to combine the terms in the fifth and sixth lines of Eq.~\eqref{eq:total-action-decomposed-caps-canceled}
with those in the last line.
To accomplish this endeavor, the following chain of steps turns out to
be serviceable \cite{Poisson:2002,Poisson:2004}:
\begin{align}
\label{eq:introduction-intrinsic-curvature-k}
\mathcal{K}-r_{\mu}n^{\nu}\nabla_{\nu}n^{\mu}&=\mathcal{K}+(\nabla_{\nu}r_{\mu})n^{\mu} n^{\nu} \notag \\
&=\nabla_{\nu}r_{\mu}(g^{\mu \nu}-r^{\mu}r^{\nu}+n^{\mu} n^{\nu}) \notag \\
&=\nabla_{\nu}r_{\mu} (\sigma^{AB}e^{\mu}_{A} e^{\nu}_{B}) \notag \\
&=\sigma^{AB}(\nabla_{\nu}r_{\mu}e^{\mu}_{A}e^{\nu}_{B}) \notag \\
&=\sigma^{AB}k_{AB}=k\,.
\end{align}
To arrive at this result, several ingredients are valuable. First, we 
benefit from the identity $r_{\mu}\nabla_{\nu}n^{\mu}=-n^{\mu}\nabla_{\nu}r_{\mu}$, 
which follows from $r^{\mu}n_{\mu}=0$. Second, we express the trace $\mathcal{K}$ 
of the extrinsic curvature on $\mathcal{B}$ in terms of the metric of 
$\mathcal{M}$ and $r_{\mu}$, which is the unit normal of $\mathcal{B}$. 
This is possible, as $\mathcal{B}$ is a timelike hypersurface embedded into~$\mathcal{M}$:
\begin{align} 
\mathcal K&=  \gamma^{ij} \mathcal K_{ij}= \gamma^{ij} (   \nabla_{\nu} r_{\mu}    e^{\mu}_i  e^{\nu}_j )
=(\nabla_{\nu} r_{\mu})\gamma^{ij}   e^{\mu}_i  e^{\nu}_j
\notag \\ &= \nabla_{\nu} r_{\mu}  (g^{\mu \nu}-  r^{\mu} r^{\nu} )\,.
\end{align}
Third, we also interpret $S_t$ as being embedded into $\mathcal{M}$, which 
implies the induced metric $\sigma_{AB}=g_{\alpha\beta} e^{\alpha}_{A}
e^{\beta}_{B}$ on $S_t$ with the valued tensors 
$e^{\alpha}_{A}:=\partial x^{\alpha}/\partial \theta^A$. 
Last but not least, due to their embedding into $\mathcal{M}$, 
the two-dimensional hypersurfaces $S_t$ also have an extrinsic 
curvature associated with them as do $\Sigma_t$ and $\mathcal{B}$. 
The latter is frequently denoted as $k_{AB}$ in the 
literature~\cite{Poisson:2002,Poisson:2004} where $k:=k_{AB}\sigma^{AB}$ is its corresponding trace.

To handle the terms in Eq.~\eqref{eq:total-action-decomposed-caps-canceled} 
on $\mathcal{B}$ containing $s^{ij}$, consider $k_{AB}:=e^i_Ae^j_Bk_{ij}$ 
with the valued tensors $e^i_A:=\partial z^i/\partial \theta^A$ based on 
the embedding of $S_t$ into $\mathcal{B}$. So,
\begin{align} 
\label{eq:relation-extrinsic-curvature-B-St}
k_{ij}&=\tilde D_ir_j=\sigma^{k}_{\phantom{k}i} \sigma^{l}_{\phantom{l}j}  D_k r_l= \sigma^{k}_{\phantom{k}i} D_kr_j
\notag \\
&=  (\delta^{k}_{\phantom{k}i}-  r_{i} r^{k} )   D_{k} r_{j}  =\mathcal K_{ij}-r_ia_j\,,
\end{align}
where $\tilde D_i$ is the covariant derivative compatible with $\sigma_{ij}$ 
and $\sigma^k_{\phantom{k}i}=\delta^k_{\phantom{k}i}-r^kr_i$ projects a 
part of a tensor living in $\Sigma_t$ onto $S_t$. This object is 
analogous to the projector $q^{\mu}_{\phantom{\mu}\nu}$ introduced in 
Eq.~\eqref{eq:decomposition-s} that is responsible for projections 
from $\mathcal{M}$ onto $\Sigma_t$. Note the sign difference in the 
second terms on the right-hand sides of the definitions of $q^{\mu}_{\phantom{\mu}\nu}$ 
and $\sigma^k_{\phantom{k}i}$, which is due to $n_{\mu}$ being timelike and 
$r_{\mu}$ being spacelike. Moreover, $r^l D_k r_l=0$ is employed in the 
first line of Eq.~\eqref{eq:relation-extrinsic-curvature-B-St}. So the latter equation
means that the extrinsic curvature $k_{AB}$ of $S_t$ lifted to 
$\mathcal{B}$ is expressed through the extrinsic curvature 
$\mathcal{K}_{ij}$ of $\mathcal{B}$.

Finally, we foliate $\mathcal{B}$ in terms of $S_t$ and recast the corresponding integral 
measure into the following form: $\mathrm{d}^3z\,\sqrt{-\gamma}=\mathrm{d}t\mathrm{d}^2\theta\,N\sqrt{\sigma}$. 
Now it makes sense to define the tensor field $s^{AB}$, which can be interpreted as $s^{ij}$ 
restricted to $S_t$. The former exists because of the embedding of $S_t$ into $\mathcal{B}$. 
We introduce $s^{AB}$ in a manner analogous to how we defined $s^{ab}$ implicitly as 
$s^{\mu\nu}$ restricted to $\Sigma_t$ via Eq.~\eqref{eq:spacelike-s}. The defining relationship
is $\sigma^i_{\phantom{i}m} \sigma^j_{\phantom{j}n} s^{mn}=:e^i_Ae^j_Bs^{AB}$ with the valued tensors defined 
directly above Eq.~\eqref{eq:relation-extrinsic-curvature-B-St}. As a consequence, 
$k_{ij}s^{ij}=k_{AB}s^{AB}$ can be deduced on $S_t$. We then arrive at the final 
form of the ADM-decomposed action, which is one of the central results of the current
work:
\begin{subequations}
\label{eq:action-G-complete}
\begin{align}\label{ActionG}
S_G&=\int_{t_1}^{t_2} \mathrm{d}t \,(L_b+B_S)\,,
\end{align}
with the Lagrangian in the bulk,
\begin{align}
\label{eq:lagrangian-bulk}
L_b&=\int_{\Sigma_t} \mathrm{d}^3y\,\mathcal{L}_b\,, \displaybreak[0]\\[2ex]
\label{eq:lagrange-density}
\mathcal{L}_b&=\frac{N\sqrt{q}}{2\kappa}\bigg[(1-u)(R-K^2+K_{ab}K^{ab})  \notag \\
&\phantom{{}={}}\hspace{1cm}+s^{\mathbf{nn}}(K^2-K_{ab}K^{ab})+s^{ab} {R}_{ab} \notag 
\\
&\phantom{{}={}}\hspace{1cm}-\frac 1NK_{ab} \mathcal{L}_ms^{ab}-2K_{ab} K^a_{\phantom{a}c} s^{cb} +a_b D_a s^{ab} \notag \\
&\phantom{{}={}}\hspace{1cm}-a^cD_c(2u+s^{\mathbf{nn}})+\frac{K}{N}(2\mathcal{L}_mu+\mathcal{L}_ms^{\mathbf{nn}})\bigg]\,,
\end{align}
and the boundary term
\begin{equation}
\label{eq:boundary-term-full}
B_S=\oint_{S_t} \mathrm{d}^2\theta\,\frac{N\sqrt{\sigma}}{2\kappa}\,\left[2(1-u)k+k_{AB}s^{AB}\right]\,.
\end{equation}
\end{subequations}
Let us summarize what we did. We ADM-decomposed the modified EH action stated in 
Eq.~\eqref{eq:action-bulk} including the extended GHY boundary term of Eq.~\eqref{Baction}. 
The latter was shown to partially cancel with boundary terms arising from total covariant 
derivatives in the bulk action. A piece of the extended GHY boundary term evaluated on the 
two-dimensional hypersurfaces $S_t$ remained. This part, which is given by Eq.~\eqref{eq:boundary-term-full},
was expressed completely in terms of quantities living in $S_t$.

Now, the resulting bulk Lagrangian of Eq.~\eqref{eq:lagrangian-bulk} involves four 
classes of terms. First, there are contributions depending on $R$ and $R_{ab}$, 
i.e., they encode the intrinsic geometry of $\Sigma_t$. Such terms occur in GR, the $u$, and 
the $s^{ab}$ sectors, but not for $s^{\mathbf{nn}}$. Second, terms quadratic in 
the extrinsic curvature or its trace are found for all sectors. Third, each sector 
comes with a Lie derivative of the corresponding SME coefficients for $m^{\mu}$. 
Last but not least, there are three contributions involving the ADM acceleration. 
Note that the surface term of Eq.~\eqref{eq:boundary-term-full} does not depend on 
the background field $s^{\mathbf{nn}}$, which is closely related to the observation 
of there being no term of the form $s^{\mathbf{nn}}R$ in $L_b$. This property 
is to be explained in more detail below.
\section{Palatini method of variation}
\label{sec:modified-dynamics}
Our recent work \cite{Reyes:2022mvm} is dedicated to a derivation of 
the modified Einstein equations based on Eq.~\eqref{eq:modified-EH} by 
resorting to the Hamiltonian formulation of this theory. Our incentive 
was to verify whether or not the Hamiltonian approach gives rise to the same 
dynamics as does the covariant formulation. The reply to this question 
was found to be in the affirmative, i.e., both approaches can be neatly connected to each other.

In the following, we intend to derive the dynamical field equations 
again, but this time by using a different approach that incorporates 
a detailed analysis of the boundary terms. Such a treatment can be beneficial in the future to explore
the limit of asymptotic flatness. Besides, as an extension of Ref.~\cite{Reyes:2022mvm},
we now allow for a nonzero shift vector $N^a$. Doing so poses a natural next step, as the
shift vector it needed to change coordinates when going from one spatial hypersurface
$\Sigma_t$ to the next.

Studying a dynamical process in numerical relativity, e.g., the frame dragging effect
of a Kerr black hole or the collapse of a star into a black hole, one finds that coordinates
can get twisted such that coordinate singularities and even physical singularities may arise.
There exists a gauge known as minimal distortion \cite{Smarr:1978dia,Gourgoulhon:2007ue,Gourgoulhon:2012}
that relies on the shift vector as a means to compensate the twisting of coordinate lines.
Thus, to be able to treat gravity systems numerically, a nonzero shift vector seems
indispensable.

Now, we will dedicate ourselves to the dynamics of the 
modified-gravity theory stated in Eq.~\eqref{eq:modified-EH}.
To do so, we consider the Palatini method of variation \cite{Misner:1973}, in which coordinate and
momentum variables are treated as independent.
In our case, the Palatini action is expressed in terms of a generic 
induced metric $q_{ab}$ and the corresponding canonical 
momentum $\Pi^{ab}$ as
\begin{align}
\label{eq:Gaction}
S_G =\int_{t_2}^{t_1} \mathrm{d}t \,   \left[\int_{\Sigma_t}  
\Pi^{ab} \dot q_{ab} \mathrm{d}^3 y -H_G(\Pi^{ab} ,q_{ab})  \right]\,,
\end{align}
where $H_G$ is the total Hamiltonian that contains a boundary 
term coming from Eq.~\eqref{eq:boundary-term-full}.
By considering 
\begin{align}
\label{eq:variation-hamiltonian}
\delta H_G  =  \int_{\Sigma_t} \mathrm{d}^3 y\,( \mathcal P^{ab} \delta q_{ab}  +\mathcal F_{ab}
\delta \Pi^{ab}   )\,,
\end{align}
a variation of the action in Eq.~\eqref{eq:Gaction} leads to
\begin{align}
\label{eq:Gaction2}
\delta S_G &=\int  _{t_1}^{t_2} \mathrm{d}t  
\int_{\Sigma_t}  \mathrm{d}^3 y  \,\Big[ \left( \dot q_{ab}- \mathcal F_{ab} \right) 
\delta \Pi^{ab} \nonumber \\
&\phantom{{}={}}\hspace{2.4cm}-\left(  {\dot {\Pi}}^{ab} +  \mathcal P^{ab}   \right)   \delta q_{ab}   \Big]\,,
\end{align}
where we have neglected a contribution that arises from an integration 
by parts, since $\delta q_{ab}=0$ on the boundary, by definition. In principle, the above variation may lead to
boundary terms depending on covariant derivatives of $\delta q_{ab}$ along the normal direction $r_c$ of $S_t$. Also, one may have
to include boundary terms already contained in the action. In a rigorous treatment, each of these contributions should be kept
track of in the derivation.

Now, the requirement that the action be stationary implies the following field 
equations in the Palatini formalism:
\begin{subequations}
\begin{align}
\label{eq:field-equation-palatini-1}
\dot q_{ab}&=   \mathcal F_{ab}    \,, \\[2ex]
\label{eq:field-equation-palatini-2}
{\dot {\Pi}}^{ab}&=- \mathcal P^{ab}     \,.
\end{align}
\end{subequations}
It is challenging to invert the extrinsic curvature for the canonical momentum, i.e.,
to compute the Hamiltonian $H_G$ when all SME coefficients $u,s^{\mathbf{nn}}$, and $s^{ab}$ are present simultaneously.
Therefore, we will be restricting ourselves to three separate analyses below, as we
already did in previous works~\cite{Reyes:2021cpx,Reyes:2022mvm,Reyes:2022dil}.

On the one hand, in each of these cases, Eq.~\eqref{eq:field-equation-palatini-1} gives 
rise to the generic geometric identity~\cite{Misner:1973,Carlip:1998,Thiemann:2007zz}
\begin{equation}
\label{eq:definition-extrinsic-curvature}
\dot{q}_{ab}=2NK_{ab}+D_aN_b+D_bN_a\,,
\end{equation}
which, in principle, corresponds to the definition of the extrinsic 
curvature. This relation remains unmodified, even in the presence of $u$ and $s^{\mu\nu}$, since 
the geometric setting is still pseudo-Riemannian geometry.

On the other hand, Eq.~\eqref{eq:field-equation-palatini-2}
encodes the dynamics of the modified-gravity theory under study. Thus, we will focus on the latter,
as it describes how gravitational dynamics is affected by diffeomorphism violation. However, if we
were working with Eq.~\eqref{eq:field-equation-palatini-2} directly, there would be no chance of taking into
account possible boundary terms. Instead, in what follows, we will cast the action of Eq.~\eqref{eq:action-G-complete}
into the form of Eq.~\eqref{eq:Gaction2} and compute the variation of the Hamiltonian $H_G$ for $q_{ab}$.
After taking proper care of boundary terms, the integral over $\Sigma_t$ can be dropped, which leads us
automatically to Eq.~\eqref{eq:field-equation-palatini-2}, evaluated for the specific sector being explored.

\subsection{Dynamics in the $u$ sector}
First, we focus on the $u$ sector, i.e., let $\mathcal{L}_u$ be 
the Lagrange density following from Eq.~\eqref{eq:lagrange-density} 
such that $\mathcal{L}_u:=\mathcal{L}_b|_{s^{\mathbf{nn}}=s^{ab}=0}$. The canonical momentum then reads
\begin{align}
\label{eq:relation-canonical-momentum-extrinsic-curvature-u}
\pi^{ab}&:=\frac{\partial\mathcal{L}_u}{\partial \dot{q}_{ab}} \notag \\
&=\frac{\sqrt{q}}{2\kappa}\left[(1-u)(K^{ab}-q^{ab}K)+\frac{1}{N}q^{ab}\mathcal{L}_mu\right]\,.
\end{align}
Recall that the extrinsic curvature proper is the standard quantity of pseudo-Riemannian 
geometry; cf.~Eq.~\eqref{eq:definition-extrinsic-curvature}. However, relationships between canonical variables
and geometrical quantities are  affected by diffeomorphism violation, which is observed here.

The ADM-decomposed action $S_{G,u}$ is
\begin{equation}
S_{G,u}=\int_{t_1}^{t_2}\mathrm{d}t\int_{\Sigma_t} \mathrm{d}^3y\,\mathcal{L}_u\,.
\end{equation}
To apply the Palatini formalism, the latter must be expressed in terms 
of a Hamiltonian via an inverse Legendre transformation:
\begin{subequations}
\begin{equation}
\label{eq:action-sector-u}
S_{G,u}=\int_{t_1}^{t_2} \mathrm{d}t \left(\int_{\Sigma_t}   \pi^{ab}\dot{q}_{ab} \, \mathrm{d}^3y-H_{\Sigma,u}+B_{S,u}\right)\,,
\end{equation}
with the bulk Hamiltonian
\begin{align}
\label{eq:bulk-hamiltonian-u}
H_{\Sigma,u}&=\int_{\Sigma_t}\mathrm{d}^3y\,\bigg\{-\frac{N\sqrt{q}}{2\kappa}\left[(1-u)R-2a^cD_cu\right] \notag \\
&\phantom{{}={}}\hspace{1.4cm}+\frac{2\kappa N}{\sqrt{q}(1-u)}\left(\pi^{ab}\pi_{ab}
-\frac{\pi^2}{2}\right) \notag \\
&\phantom{{}={}}\hspace{1.4cm}+\frac{\mathcal{L}_mu}{1-u}\left(\pi-\frac{3}{4}
\frac{\sqrt{q}}{\kappa N}\mathcal{L}_mu\right) \notag \\
&\phantom{{}={}}\hspace{1.4cm}+2\pi^{ab}D_aN_b\bigg\}\,,
\end{align}
and the boundary term of Eq.~\eqref{eq:boundary-term-full} restricted to $u$:
\begin{equation}
\label{eq:boundary-term-lagrangian-u}
B_{S,u}=\oint_{S_t}\mathrm{d}^2\theta\,\frac{N\sqrt{\sigma}}{\kappa}(1-u)k\,.
\end{equation}
\end{subequations}
In the following, we intend to evaluate the variations for $q_{ij}$ of 
each term in $H_{\Sigma,u}$ that contributes to the action of Eq.~\eqref{eq:action-sector-u}. 
First, the variation of the Lie derivative of $u$ with respect to $m^{\mu}$ is given by
\begin{align}
\label{eq:variation-lie-derivative-u}
\delta\int_{\Sigma_t}\mathrm{d}^3y\,\mathcal{L}_mu&=
\delta\int_{\Sigma_t}\mathrm{d}^3y\,(\dot{u}-\mathcal{L}_Nu) \notag \\
&=\int_{\Sigma_t}\mathrm{d}^3y\,\delta(\dot{u}-q_{cd}N^cD^du) \notag \\
&=\int_{\Sigma_t}\mathrm{d}^3y\,(-N^{(a}D^{b)}u)\delta q_{ab}\,,
\end{align}
where we denote the symmetrization of tensors by pairs of 
parentheses around indices, i.e., $X^{(a}Y^{b)}:=(X^aY^b+X^bY^a)/2$. So this variation is 
nonzero only when coordinates with a nonvanishing shift vector are considered. 

Next, the variation of the term including the Ricci tensor requires an integration by
parts generating a nonvanishing boundary term on $S_t$. We
proceed to present the calculation with some detail. Although the 
latter bears many similarities with the corresponding computation
done in GR, its exposition is still expected to be worthwhile for the reader, as 
it may serve as a foundation to understand the more intricate analysis 
in the $s^{ab}$ sector to be done later. 
As a warm-up, it is useful to consider the variation
\begin{align}
\delta( \sqrt q R)&=\Big(-\sqrt{q} G^{ab}+D^aD^bN-q^{ab}D_cD^cN\Big)\delta q_{ab} \notag \\
&\phantom{{}={}}+\sqrt{q}D_c \delta V^c\,,
\end{align}
where we have defined the contravariant Einstein tensor on $\Sigma_t$ by $G^{ab}:=R^{ab}-(R/2)q^{ab}$
and used that $\delta q_{ab}$ vanishes on $S_t$ in a second integration by parts. 
Moreover, we introduced the quantity
\begin{equation}
\delta V^c=q^{ab}\delta\Gamma^c_{\phantom{c}ab}- q^{ac}\delta\Gamma^b_{\phantom{b}ab}\,,
\end{equation}
which includes variations of the Christoffel symbols of~$\Sigma_t$. They can be 
expressed in terms of variations of the corresponding intrinsic metric:
\begin{align}
\delta\Gamma^c_{\phantom{c}ab}=\frac{1}{2}q^{cf}(D_a\delta q_{fb}+D_b\delta q_{fa}-D_f\delta q_{ab})\,.
\end{align}
Applying these ingredients to the $u$ sector leads to
\begin{align}
\delta\int_{\Sigma_t}&\mathrm{d}^3y\,\frac{N\sqrt{q}}{2\kappa}(1-u)R \notag \\
&=\int_{\Sigma_t}\mathrm{d}^3y\,\frac{\sqrt{q}}
{2\kappa}\Big(-N(1-u)G^{ab}+D^aD^b[(1-u)N] \notag \displaybreak[0]\\[-1ex]
&\phantom{{}={}}\hspace{2cm} -q^{ab}D_c
D^c[(1-u)N]\Big)\delta q_{ab}\notag \\
&\phantom{{}={}}+\int_{\Sigma_t} \mathrm{d}^3y\,\frac{\sqrt{q}}{2\kappa}
\,D_c[N (1-u) \delta V^c]\,.
\end{align}
The last integral on the right-hand side can be evaluated with Gauss' theorem to provide
\begin{align}
\int_{\Sigma_t} \mathrm{d}^3y\,&\frac{\sqrt{q}}{2\kappa}
\,D_c[N (1-u) \delta V^c] \notag \\
&=\oint_{S_t} \mathrm{d}^2\theta \,\frac{N\sqrt{\sigma}}{2\kappa}
\,(1-u) r_c \delta V^c\,.
\end{align}
By using 
\begin{align}
r_c \delta V^c =-\sigma^{ab} r^c D_c \delta q_{ab}\,,
\end{align}
we arrive at 
\begin{subequations}
\begin{align}
\label{eq:variation-ricci-with-boundary-term-u}
\delta\int_{\Sigma_t}&\mathrm{d}^3y\,\frac{N\sqrt{q}}{2\kappa}(1-u)R \notag \\
&=\int_{\Sigma_t}\mathrm{d}^3y\,\frac{\sqrt{q}}
{2\kappa}\Big(-N(1-u)G^{ab}+D^aD^b[(1-u)N] \notag \displaybreak[0]\\[-1ex]
&\phantom{{}={}}\hspace{1.6cm} -q^{ab}D_c
D^c[(1-u)N]\Big)\delta q_{ab}-\delta B_{R,u}\,,
\end{align}
with a boundary term denoted as $\delta B_{R,u}$, which reads
\begin{equation}
\label{eq:variation-boundary-ricci-u}
\delta B_{R,u}=\oint_{S_t} \mathrm{d}^2\theta\,
\frac{N\sqrt{\sigma}}{2\kappa}(1-u)\sigma^{ab} r^cD_c \delta q_{ab}\,.
\end{equation}
\end{subequations}
We continue with the remaining variations necessary:
\begin{align}
\label{eq:variation-shift-vector-term}
\delta\int_{\Sigma_t}\mathrm{d}^3y\,&\pi^{cd}D_cN_d \notag \\
&=\int_{\Sigma_t}\mathrm{d}^3y\,\frac{\sqrt{q}}{2}
D_c\left(\frac{2N^{(a}\pi^{b)c}-\pi^{ab}N^c}{\sqrt{q}}\right)\delta q_{ab}\,,
\end{align}
as well as
\begin{widetext}
\begin{subequations}
\label{eq:variations-momentum-and-lie-derivative-terms-u}
\begin{align}
\delta\int_{\Sigma_t}\mathrm{d}^3y\,\frac{2\kappa N}{\sqrt{q}(1-u)}\left(\pi^{cd}\pi_{cd}-
\frac{\pi^2}{2}\right)&=\int_{\Sigma_t}\mathrm{d}^3y\,\frac{2\kappa N}
{\sqrt{q}(1-u)}\bigg[2\pi^{ac}\pi^b_{\phantom{a}c}-\pi\pi^{ab}-\frac{1}{2}
\left(\pi_{cd}\pi^{cd}-\frac{\pi^2}{2}\right)q^{ab}\bigg]\delta q_{ab}\,, \displaybreak[0]\\[2ex]
\delta\int_{\Sigma_t}\mathrm{d}^3y\,\frac{\mathcal{L}_mu}
{1-u}\left(\pi-\frac{3}{4}\frac{\sqrt{q}}{\kappa N}\mathcal{L}_mu\right) &=
\int_{\Sigma_t}\mathrm{d}^3y\,\bigg[\frac{\mathcal{L}_mu}{1-u}\left(\pi^{ab}
-\frac{3}{8}\frac{\sqrt{q}}{\kappa N}q^{ab}\mathcal{L}_mu\right)
-\frac{N^{(a}D^{b)}u}{1-u}\left(\pi-\frac{3}{2}\frac{\sqrt{q}}{\kappa N}\mathcal{L}_mu\right)\bigg]\delta q_{ab}\,.
\end{align}
\end{subequations}
\end{widetext}
Finally, the variation of the boundary term in the action, Eq.~\eqref{eq:boundary-term-lagrangian-u}, remains to be computed:
\begin{subequations}
\label{eq:variation-boundary-term-action-u}
\begin{equation}
\delta B_{S,u}=\oint_{S_t} \mathrm{d}^2\theta\,\frac{N\sqrt{\sigma}}{\kappa}(1-u) \delta k\,,
\end{equation}
with the variation of the extrinsic-curvature scalar $k$ on $S_t$, which can be expressed as
\begin{equation}
\label{expK}
\delta k=\delta (\sigma^{ab} D_a r_b)=\frac 12 \sigma^{ab} r^cD_c\delta q_{ab}\,.
\end{equation}
\end{subequations}
We see that the latter cancels the boundary term of Eq.~\eqref{eq:variation-boundary-ricci-u}, 
which results from varying the contribution proportional to the Ricci scalar: $\delta B_{S,u}-\delta B_{R,u}=0$.

Note that a peculiar observation is made when varying the term in Eq.~\eqref{eq:bulk-hamiltonian-u} 
depending on the ADM acceleration. The variation provides a contribution to the field equations that
makes them slightly differ from the modified Einstein equations for Eq.~\eqref{eq:modified-EH},
projected onto $\Sigma_t$ (see also Ref.~\cite{Reyes:2022mvm}). The reason for this issue 
is that a term with a single covariant derivative of a background field does not occur 
in the modified covariant action of this theory, Eq.~\eqref{eq:modified-EH}, which was the starting point to obtain 
the field equations. Also, the modified Einstein equations on $\mathcal{M}$, which 
were derived for the first time in Ref.~\cite{Bailey:2006fd}, only involve second-order 
covariant derivatives of the SME background fields (see App.~V of Ref.~\cite{Reyes:2021cpx} for details on the
derivation of these equations). Furthermore, boundary terms were neglected in these derivations, in particular, contributions
proportional to the ADM acceleration, which would be an implication from varying the Ricci scalar $R$.

To remedy the mismatch mentioned, we bring our  action into a more suitable form by adding an integral
over a total derivative. The latter  gives rise to another boundary term on $S_t$:
\begin{equation}
\label{eq:boundary-term-u-matching}
B_{a,u}=\int_{\Sigma_t} \mathrm{d}^3y\,\frac{\sqrt{q}}{\kappa}D_c(uNa^c)
=\oint_{S_t} \mathrm{d}^2\theta\,\frac{N\sqrt{\sigma}}{\kappa}ur_ia^i\,,
\end{equation}
which explicitly depends on the ADM acceleration. In specific coordinates that are 
characterized by a vanishing ADM acceleration such as Gaussian normal 
coordinates~\cite{Misner:1973}, this boundary term vanishes identically and the 
issue is absent. Adding Eqs.~\eqref{eq:action-sector-u} and \eqref{eq:boundary-term-u-matching} 
makes the covariant derivative act on $a^c$, i.e., the action does then no longer involve a 
first-order covariant derivative of $u$: 
\begin{equation}
\label{eq:modifying-acceleration-term-u}
\int_{\Sigma_t}\mathrm{d}^3y\,\frac{N\sqrt{q}}{\kappa}(-a^cD_cu)+B_{a,u}=\int_{\Sigma_t}\mathrm{d}^3y\,\frac{\sqrt{q}}{\kappa}D_c(Na^c)u\,.
\end{equation}
So the variation of the total derivative in 
Eq.~\eqref{eq:boundary-term-u-matching} provides a nonvanishing contribution. Now, the variation of Eq.~\eqref{eq:modifying-acceleration-term-u} reads
\begin{align}
\label{eq:variation-adapted-acceleration-terms-u}
\delta\int_{\Sigma_t}\mathrm{d}^3y\,\frac{\sqrt{q}}{\kappa}
D_c(Na^c)u&=\int_{\Sigma_t}\mathrm{d}^3y\,\frac{N\sqrt{q}}{2\kappa}(-q^{ab}a_cD^cu \notag \displaybreak[0]\\
&\phantom{{}={}}\hspace{1.8cm}+2a^{(a}D^{b)}u)\delta q_{ab}\,.
\end{align}
After canceling the boundary terms, putting together the individual pieces of Eqs.~\eqref{eq:variation-ricci-with-boundary-term-u} (with $\delta B_{R,u}$ discarded), \eqref{eq:variation-shift-vector-term}, \eqref{eq:variations-momentum-and-lie-derivative-terms-u}, and \eqref{eq:variation-adapted-acceleration-terms-u} and inserting those into the second line of Eq.~\eqref{eq:Gaction2} leads to an integral of a second-rank tensor over $\Sigma_t$,
which must be equal to zero. The foliation and, therefore, $\Sigma_t$ is arbitrary, so is $\delta q_{ab}$.
Thus, the integral is equal to zero if and only if the integrand vanishes.
This line of reasoning implies the field equations:
\begin{widetext}
\begin{align}
\label{eq:einstein-equations-u}
\dot{\pi}^{ab}&=\frac{\sqrt{q}}{2\kappa}\Big(-N(1-u)G^{ab}+D^aD^b[(1-u)N]-q^{ab}D_cD^c[(1-u)N]\Big) \notag \\
&\phantom{{}={}}-\frac{2\kappa N}{\sqrt{q}(1-u)}\bigg[2\pi^{ac}
\pi^b_{\phantom{b}c}-\pi\pi^{ab}-\frac{1}{2}\left(\pi_{cd}\pi^{cd}
-\frac{\pi^2}{2}\right)q^{ab}\bigg]+\frac{N\sqrt{q}}{2\kappa}(2a^{(a}D^{b)}u-q^{ab}a_cD^cu) \notag \\
&\phantom{{}={}}-\frac{\mathcal{L}_mu}{1-u}\left(\pi^{ab}
-\frac{3}{8}\frac{\sqrt{q}}{\kappa N}q^{ab}\mathcal{L}_mu\right)
+\frac{N^{(a}D^{b)}u}{1-u}\left(\pi-\frac{3}{2}\frac{\sqrt{q}}{\kappa N}
\mathcal{L}_mu\right)-\sqrt{q}D_c\left(\frac{2N^{(a}\pi^{b)c}-\pi^{ab}N^c}{\sqrt{q}}\right)\,.
\end{align}
\end{widetext}
The reader can check that the latter correctly reduce to 
$(\vec{\boldsymbol{q}}^{*}\mathbf{Q})^{ij}=0$ with Eq.~(30a) 
in Ref.~\cite{Reyes:2022mvm} in the limit of $N^a=0$. These field equations 
are the physical part of the modified Einstein equations on $\mathcal{M}$ \cite{Bailey:2006fd} and
encode the dynamical information of the theory described by Eq.~\eqref{eq:modified-EH}
with $s^{\mu\nu}=0$. Modifications of the Hamiltonian and momentum constraints from GR have been
separated from the latter.
\subsection{Dynamics in the $s^{\mathbf{nn}}$ sector}
The computations are similar for the $s^{\mathbf{nn}}$ sector. We define 
the Lagrange density from Eq.~\eqref{eq:lagrange-density} by $\mathcal{L}_1:=
\mathcal{L}_b|_{u=s^{ab}=0}$. The canonical momentum follows from the latter as before:
\begin{align}
\label{eq:relation-canonical-momentum-extrinsic-curvature-snn}
p^{ab}&:=\frac{\partial\mathcal{L}_1}{\partial \dot{q}_{ab}} \notag \\
&=\frac{\sqrt{q}}{2\kappa}\left[(1-s^{\mathbf{nn}})(K^{ab}-q^{ab}K)
+\frac{1}{2N}q^{ab}\mathcal{L}_ms^{\mathbf{nn}}\right]\,,
\end{align}
i.e., the relationship between the canonical momentum and the extrinsic curvature is 
modified in a way quite similar to Eq.~\eqref{eq:relation-canonical-momentum-extrinsic-curvature-u} 
for $u$. Then, the ADM-decomposed action is
\begin{equation}
S_{G,1}=\int_{t_1}^{t_2} \mathrm{d}t\int_{\Sigma_t} \mathrm{d}^3y\,\mathcal{L}_1\,.
\end{equation}
To apply the Palatini formalism, the latter is cast into the following more suitable form:
\begin{subequations}
\begin{equation}
\label{eq:action-sector-snn}
S_{G,1}=\int_{t_1}^{t_2} \mathrm{d}t\bigg(\int_{\Sigma_t} p^{ab}\dot{q}_{ab}\, \mathrm{d}^3y -H_{\Sigma,1}+B_S\bigg)\,,
\end{equation}
with the Hamiltonian in the bulk,
\begin{align}
\label{eq:bulk-hamiltonian-snn}
H_{\Sigma,1}&=\int_{\Sigma_t} \mathrm{d}^3y\,\bigg\{-\frac{N\sqrt{q}}{2\kappa}(R-a^cD_cs^{\mathbf{nn}}) \notag \displaybreak[0]\\
&\phantom{{}={}}\hspace{1.4cm}+\frac{2\kappa N}{\sqrt{q}(1-s^{\mathbf{nn}})}\left(p^{ab}p_{ab}-\frac{p^2}{2}\right) \notag \displaybreak[0]\\
&\phantom{{}={}}\hspace{1.4cm}+\frac{\mathcal{L}_ms^{\mathbf{nn}}}{2(1-s^{\mathbf{nn}})}\left(p-\frac{3}{8}\frac{\sqrt{q}}{\kappa N}\mathcal{L}_ms^{\mathbf{nn}}\right) \notag \\
&\phantom{{}={}}\hspace{1.4cm}+2p^{ab}D_aN_b\bigg\}\,,
\end{align}
and the boundary term
\begin{equation}
\label{eq:boundary-term-lagrangian-snn}
B_S=\oint_{S_t} \mathrm{d}^2\theta\,\frac{N\sqrt{\sigma}}{\kappa}k\,,
\end{equation}
\end{subequations}
corresponding to that of GR.
Computing the variations works such as it does for $u$. It is even 
a bit simpler, since a term that multiplies the curvature scalar $R$ with 
$s^{\mathbf{nn}}$ is absent. From the point of view established until this 
moment, this property makes perfect sense. The boundary term of 
Eq.~\eqref{eq:boundary-term-lagrangian-snn} does not contain a 
piece proportional to $s^{\mathbf{nn}}$, which would have to be canceled 
against the boundary term arising from the variation of $R$, 
cf.~Eqs.~\eqref{eq:variation-boundary-ricci-u}, \eqref{eq:variation-boundary-term-action-u} for the $u$ sector.

As for the contribution in Eq.~\eqref{eq:bulk-hamiltonian-snn} depending 
on the ADM acceleration, the same hurdle that we already observed for the $u$ sector arises here, too.
The variation of this term for the induced metric leads to a deviation of the modified field 
equations from the modified Einstein equations of Ref.~\cite{Bailey:2006fd}
projected onto $\Sigma_t$. Hence, it is indispensable to consider another total-derivative correction
term:
\begin{equation}
\label{eq:boundary-term-snn-matching}
B_{a,s}=\int_{\Sigma_t} \mathrm{d}^3y\,\frac{\sqrt{q}}{2\kappa}D_c(s^{\mathbf{nn}}Na^c)=\oint_{S_t} \mathrm{d}^2\theta\,\frac{N\sqrt{\sigma}}{2\kappa}s^{\mathbf{nn}}r_ia^i\,,
\end{equation}
cf.~Eq.~\eqref{eq:boundary-term-u-matching}.
By adding the latter to Eq.~\eqref{eq:action-sector-snn}, the covariant 
derivative is moved to $a^c$, cf.~Eq.~\eqref{eq:modifying-acceleration-term-u} 
for the $u$ sector. The variation of the resulting term for the induced metric then implies
\begin{align}
\label{eq:variation-adapted-acceleration-terms-snn}
\delta\int_{\Sigma_t}\mathrm{d}^3y\,\frac{\sqrt{q}}
{2\kappa}D_c(Na^c)s^{\mathbf{nn}}&=\int_{\Sigma_t}\mathrm{d}^3y\frac{N\sqrt{q}}{4\kappa}(-q^{ab}a_cD^cs^{\mathbf{nn}} \notag \\
&\phantom{{}={}}\hspace{1.2cm}+2a^{(a}D^{b)}s^{\mathbf{nn}})\delta q_{ab}\,.
\end{align}
The remaining variations can be computed in a manner analogous to how 
we did it in Eqs.~\eqref{eq:variation-shift-vector-term}, \eqref{eq:variations-momentum-and-lie-derivative-terms-u} for $u$.
The Ricci scalar term does not involve the coefficient $s^{\mathbf{nn}}$.
Therefore, the variation of this term gives rise to the same boundary 
term on $S_t$ that must also be considered for the EH action. We will denote the latter as $-\delta B_R$. The total variation then reads
\begin{subequations}
\begin{align}
\label{eq:variation-ricci-with-boundary-term}
\delta\int_{\Sigma_t}&\mathrm{d}^3y\,\frac{N\sqrt{q}}{2\kappa}R \notag \\
&=\int_{\Sigma_t}\mathrm{d}^3y\,\frac{\sqrt{q}}{2\kappa}(D^aD^bN-q^{ab}D_cD^cN \notag \\[-1ex]
&\phantom{{}={}}\hspace{1.8cm}-NG^{ab})\delta q_{ab}-\delta B_R\,,
\end{align}
with
\begin{equation}
\label{eq:variation-boundary-ricci-snn}
\delta B_R=\oint_{S_t} \mathrm{d}^2\theta\,\frac{N\sqrt{\sigma}}{2\kappa}\sigma^{ab}r^cD_c\delta q_{ab}\,.
\end{equation}
\end{subequations}
Here, we have used the result of Eq.~\eqref{eq:variation-boundary-ricci-u} for $u=0$. 
Considering the variation of the boundary term in the action, 
i.e., Eq.~\eqref{eq:boundary-term-lagrangian-snn}, leads to
\begin{equation}
\label{eq:variation-boundary-action-snn}
\delta B_S=\oint_{S_t} \mathrm{d}^2\theta\,\frac{N\sqrt{\sigma}}{\kappa} \delta k\,.
\end{equation}
By benefiting from Eq.~\eqref{expK}, the boundary term of Eq.~\eqref{eq:variation-boundary-ricci-snn}
compensates the variation of Eq.~\eqref{eq:variation-boundary-action-snn}, as expected: $\delta B_S-\delta B_R=0$.

Furthermore,
\begin{subequations}
\label{eq:variations-momentum-and-lie-derivative-terms-snn}
\begin{align}
\delta\int_{\Sigma_t} 
&\mathrm{d}^3y\,\frac{2\kappa N}{\sqrt{q}(1-s^{\mathbf{nn}})}
\left(p^{cd}p_{cd}-\frac{p^2}{2}\right)  \notag \\ 
&=\int_{\Sigma_t}\mathrm{d}^3y\,\frac{2\kappa N}{\sqrt{q}(1-s^{\mathbf{nn}})}\bigg[2p^{ac}p^b_{\phantom{b}c}-pp^{ab} \notag \\
&\phantom{{}={}}\hspace{2.8cm}-\frac{1}{2}\left(p_{cd}p^{cd}-\frac{p^2}{2}\right)q^{ab}\bigg]\delta q_{ab}\,, \displaybreak[0] \\[2ex]
\delta\int_{\Sigma_t}&\mathrm{d}^3y\,\frac{\mathcal{L}_ms^{\mathbf{nn}}}{2(1-s^{\mathbf{nn}})}\left(p-\frac{3}{8}\frac{\sqrt{q}}{\kappa N}\mathcal{L}_ms^{\mathbf{nn}}\right)\notag  \\
&=\int_{\Sigma_t}\mathrm{d}^3y\,\bigg[\frac{\mathcal{L}_ms^{\mathbf{nn}}}{2(1-s^{\mathbf{nn}})}\left(p^{ab}-\frac{3}{16}\frac{\sqrt{q}}{\kappa N}q^{ab}\mathcal{L}_ms^{\mathbf{nn}}\right) \notag \\
&\phantom{{}={}}\hspace{1.0cm}-\frac{N^{(a}D^{b)}s^{\mathbf{nn}}}{2(1-s^{\mathbf{nn}})}\left(p-\frac{3}{4}\frac{\sqrt{q}}{\kappa N}\mathcal{L}_ms^{\mathbf{nn}}\right)\bigg]\delta q_{ab}\,.
\end{align}
\end{subequations}
Compiling the variations of Eqs.~\eqref{eq:variation-adapted-acceleration-terms-snn}, \eqref{eq:variation-ricci-with-boundary-term} (with $\delta B_R$ dropped), \eqref{eq:variations-momentum-and-lie-derivative-terms-snn} as well as the analog of Eq.~\eqref{eq:variation-shift-vector-term} and inserting them into the second line of Eq.~\eqref{eq:Gaction2} implies another second-rank tensor integrated over $\Sigma_t$, which has to vanish. The same argument that we employed for $u$ previously results in the dynamical part of the modified Einstein equations:
\begin{align}
\label{eq:einstein-equations-snn}
\dot{p}^{ab}&=\frac{\sqrt{q}}{2\kappa}\left(-NG^{ab}+D^aD^bN-q^{ab}D_cD^cN\right) \notag \displaybreak[0]\\
&\phantom{{}={}}-\frac{2\kappa N}{\sqrt{q}(1-s^{\mathbf{nn}})}\bigg[2p^{ac}p^b_{\phantom{b}c}-pp^{ab} \notag \displaybreak[0]\\
&\phantom{{}={}}\hspace{2.5cm}-\frac{1}{2}\left(p_{cd}p^{cd}-\frac{p^2}{2}\right)q^{ab}\bigg] \notag \displaybreak[0]\\
&\phantom{{}={}}+\frac{N\sqrt{q}}{4\kappa}(2a^{(a}D^{b)}s^{\mathbf{nn}}-q^{ab}a_cD^cs^{\mathbf{nn}}) \notag \displaybreak[0]\\
&\phantom{{}={}}-\frac{\mathcal{L}_ms^{\mathbf{nn}}}{2(1-s^{\mathbf{nn}})}\left(p^{ab}-\frac{3}{16}\frac{\sqrt{q}}{\kappa N}q^{ab}\mathcal{L}_ms^{\mathbf{nn}}\right) \notag \displaybreak[0]\\
&\phantom{{}={}}+\frac{N^{(a}D^{b)}s^{\mathbf{nn}}}{2(1-s^{\mathbf{nn}})}\left(p-\frac{3}{4}\frac{\sqrt{q}}{\kappa N}\mathcal{L}_ms^{\mathbf{nn}}\right) \notag \displaybreak[0]\\
&\phantom{{}={}}-\sqrt{q}D_c\left(\frac{2N^{(a}p^{b)c}-p^{ab}N^c}{\sqrt{q}}\right)\,.
\end{align}
The validity of $(\vec{\boldsymbol{q}}^{*}\mathbf{J}_1)^{ij}=0$ based on Eq.~(35a) in Ref.~\cite{Reyes:2022mvm} is confirmed for $N^a=0$. Similarly, Eq.~\eqref{eq:einstein-equations-snn} describes the dynamics of the modified-gravity theory governed by $\mathcal{L}_1$.
\subsection{Dynamics in the $s^{ab}$ sector}\label{sec:dynamics-sab-sector}
Last but not least, let $\mathcal{L}_2$ be the Lagrange 
density based on Eq.~\eqref{eq:lagrange-density} restricted 
to a nonzero $s^{ab}$ only, i.e., $\mathcal{L}_2:=\mathcal{L}_b|_{u=s^{\mathbf{nn}}=0}$. 
Then, the canonical momentum is given by
\begin{align}
\label{eq:relation-canonical-momentum-extrinsic-curvature-sab}
{P}^{ab}&:=\frac{\partial\mathcal{L}_2}{\partial\dot{q}_{ab}}  \\
&=\frac{\sqrt{q}}{2\kappa}\bigg[K^{ab}-q^{ab}K-(s^{ac}K_c^{\phantom{c}b}+s^{bc}K_c^{\phantom{c}a}) \notag \\
&\phantom{{}={}}\hspace{0.7cm}-\frac{1}{2N}\mathcal{L}_ms^{ab}\bigg] \notag  \,.
\end{align}
Due to the tensorial nature of $s^{ab}$, the latter relation has a more complicated structure as did Eqs.~\eqref{eq:relation-canonical-momentum-extrinsic-curvature-u}, \eqref{eq:relation-canonical-momentum-extrinsic-curvature-snn} for $u$ and $s^{\mathbf{nn}}$, respectively. Therefore, it does not come as a surprise that the $s^{ab}$ sector is involved from a calculational perspective. After all, it involves
six independent coefficients, which makes it challenging to invert Eq.~\eqref{eq:relation-canonical-momentum-extrinsic-curvature-sab} for the extrinsic curvature in a closed form. Therefore, as we did before in Refs.~\cite{Reyes:2021cpx,Reyes:2022mvm}, we will be working at 
first order in $s^{ab}$ and derivatives thereof.

Applying the $(3+1)$ decomposition to the action then implies
\begin{equation}
S_{G,2}=\int_{t_1}^{t_2} \mathrm{d}t\int_{\Sigma_t} \mathrm{d}^3y\,\mathcal{L}_2\,.
\end{equation}
Again, the latter is expressed in a form adequate for the Palatini formalism:
\begin{subequations}
\begin{equation}
\label{eq:action-sector-sij}
S_{G,2}=\int_{t_1}^{t_2} \mathrm{d}t \bigg(\int_{\Sigma_t} P^{ab}\dot{q}_{ab} \, \mathrm{d}^3y -H_{\Sigma,2}+B_{S,2}\bigg)\,,
\end{equation}
with the bulk Hamiltonian
\begin{align}
\label{eq:bulk-hamiltonian-sab}
H_{\Sigma,2}&=\int_{\Sigma_t}\mathrm{d}^3y\,\bigg\{-\frac{N\sqrt{q}}{2\kappa}\left(R+s^{ab}R_{ab}+a_bD_as^{ab}\right) \notag \\
&\phantom{{}={}}\hspace{1.4cm}+\left(P_{ab}-\frac{P}{2}q_{ab}\right)\mathcal{L}_ms^{ab} \notag \\
&\phantom{{}={}}\hspace{1.4cm}+\frac{2\kappa N}{\sqrt{q}}\bigg[P^{ab}P_{ab}-(1-s^a_{\phantom{a}a})\frac{P^2}{2}\notag \\
&\phantom{{}={}}\hspace{2.8cm}-2s^{ab}(P_{ab}P-P_a^{\phantom{a}c}P_{cb})\bigg] \notag \\
&\phantom{{}={}}\hspace{1.4cm}+2P^{ab}D_aN_b\bigg\}\,,
\end{align}
and the boundary term
\begin{equation}
\label{eq:boundary-term-lagrangian-sab}
B_{S,2}=\oint_{S_t} \mathrm{d}^2\theta\,\frac{N\sqrt{\sigma}}{2\kappa}\,(2k+k_{AB}s^{AB})\,.
\end{equation}
\end{subequations}
Note that $S_{G,2}$ does not require an extra total-derivative correction term to match the dynamical field equations with the modified Einstein equations on $\mathcal{M}$ \cite{Bailey:2006fd}, projected onto $\Sigma_t$. This property is in contrast to what we found for the $u$ and $s^{\mathbf{nn}}$ sectors; cf.~Eqs.~\eqref{eq:boundary-term-u-matching} and \eqref{eq:boundary-term-snn-matching}, respectively. We will come back to that point later.

Obtaining the dynamical field equations through the variation of the action is tedious. However, the computational steps involved are similar to those of the $u$ and $s^{\mathbf{nn}}$ sectors investigated before. An additional ingredient necessary to analyze is the Lie derivative of the tensor-valued background field $s^{ab}$ with respect to $m^{\mu}$, which is expressed in terms of partial derivatives as \cite{Carroll:1997ar}
\begin{align}
\mathcal{L}_ms^{cd}&=\dot{s}^{cd}-N^e\partial_es^{cd}-(\partial_fm^c)s^{fd}-(\partial_fm^d)s^{cf} \notag \\
&=\dot{s}^{cd}-q_{ef}N^e\partial^fs^{cd} \notag \\
&\phantom{{}={}}+q_{fg}(\partial^gN^c)s^{fd}+q_{fg}(\partial^gN^d)s^{cf}\,.
\end{align}
Its variation for the induced metric on $\Sigma_t$ amounts to
\begin{align}
\delta\int_{\Sigma_t} \mathrm{d}^3y\,\mathcal{L}_ms^{cd}&=\int_{\Sigma_t} \mathrm{d}^3y\,\Big[-N^{(a}\partial^{b)}s^{cd}+(\partial^{(a}N^c)s^{b)d} \notag \\
&\phantom{{}={}}\hspace{1.3cm}+(\partial^{(a}N^d)s^{b)c}\Big]\delta q_{ab}\,.
\end{align}
We also need the variation of the Ricci scalar contribution, which was already obtained in Eq.~\eqref{eq:variation-ricci-with-boundary-term}.
In a manner analogous to how the latter implies a 
boundary term on $S_t$ --- recall Eqs.~\eqref{eq:variation-boundary-ricci-u}, 
\eqref{eq:variation-boundary-ricci-snn} --- the variation of the contribution in Eq.~\eqref{eq:bulk-hamiltonian-sab}
that involves the Ricci tensor provides another boundary term on $S_t$.
To compute this variation, it is convenient to consider
\begin{subequations}
\begin{align}
\delta( \sqrt q s^{ab} R_{ab})&=\frac12 \sqrt {q} q^{cd}\delta q_{cd} R_{ab}+\sqrt{q} D_c \delta Q^c\notag \\
&\phantom{{}={}}-D_cs^{ab}  \delta \Gamma^c_{\phantom{c}ab} +D_bs^{ab}  \delta \Gamma^c_{\phantom{c}ac}\,,
\end{align}
with
\begin{align}
\delta Q^c= s^{ab}  \delta \Gamma^c_{\phantom{c}ab} -s^{ac}  \delta \Gamma^b_{\phantom{b}ab}\,.
\end{align}
We also consult the following expression in 3 dimensions analogous to Eq.~(D13) of Ref.~\cite{Reyes:2021cpx}, which is 
\begin{align}
r_c \delta Q^c= -  \frac 12    \sigma^{d}_{\phantom{d}a}  \sigma^{e}_{\phantom{e}b}  s^{ab} r^c D_c \delta q _{de}  \,,
\end{align}
where we discarded the term proportional to $s^{\mathbf{nn}}$, as the latter coefficient would have to be replaced by $s^{\mathbf{rr}}=0$ in this case.
By doing so, we arrive at the variation
\end{subequations}
\begin{subequations}
\begin{align}
\label{eq:variation-ricci-with-boundary-term-sab}
&\delta\int_{\Sigma_t}\mathrm{d}^3y\,\frac{N\sqrt{q}}{2\kappa}s^{cd}R_{cd} \notag \\
&=\int_{\Sigma_t}\mathrm{d}^3y\,\frac{N\sqrt{q}}{4\kappa}\bigg\{q^{ab}\left[R_{cd}-(D_c+a_c)(D_d+a_d)\right]s^{cd} \notag \\
&\phantom{{}={}}\hspace{1.2cm}+(D_c+a_c)\Big[(D^a+a^a)s^{bc}+(D^b+a^b)s^{ac} \notag \\
&\phantom{{}={}}\hspace{1,2cm}-(D^c+a^c)s^{ab}\Big]\bigg\}\delta q_{ab}-\delta B_{sR}\,, 
\end{align}
where
\begin{equation}
\label{eq:variation-boundary-ricci-sab}
\delta B_{sR}=\oint_{S_t} \mathrm{d}^2\theta\,\frac{N\sqrt{\sigma}}{2\kappa}\sigma^d_{\phantom{d}a} \sigma^e_{\phantom{e}b}  s^{ab} r^c D_c\delta q_{de}\,.
\end{equation}
\end{subequations}
Note that \textit{xTensor} is powerful when it comes to computing results like Eq.~\eqref{eq:variation-ricci-with-boundary-term-sab}, but it omits boundary terms such as that stated in Eq.~\eqref{eq:variation-boundary-ricci-sab}. Hence, these must be taken into account by hand.
Now, the total boundary term corresponds to the sum of Eq.~\eqref{eq:variation-boundary-ricci-snn}, which results from the variation of the EH term, and of Eq.~\eqref{eq:variation-boundary-ricci-sab}, which we have just obtained. So we define
\begin{equation}
\label{eq:variation-total-boundary-term-sab}
\delta B_{R,2}:=\delta B_R+\delta B_{sR}\,.
\end{equation}
Note also the compelling form of Eq.~\eqref{eq:variation-ricci-with-boundary-term-sab} that depends only on combinations of the covariant derivative and the ADM acceleration, $D_c+a_c$. Moreover, the variation of the term involving the ADM acceleration can also be cast into an appealing form as follows:
\begin{subequations}
\label{eq:variation-acceleration-terms-sab}
\begin{align}
\delta\int_{\Sigma_t}&\mathrm{d}^3y\,\frac{N\sqrt{q}}{2\kappa}a_cD_ds^{cd} \notag \\
&=\int_{\Sigma_t}\mathrm{d}^3y\,\bigg(-D_c\Upsilon^{abc}+\frac{N\sqrt{q}}{4\kappa}q^{ab}a_cD_ds^{cd}\bigg)\delta q_{ab}\,,
\end{align}
with
\begin{equation}
\label{eq:tensor-upsilon}
\Upsilon^{abc}=\frac{N\sqrt{q}}{4\kappa}\bigg(2a^{(a}s^{b)c}-s^{ab}a^c+q^{ab}s^{dc}a_d\bigg)\,.
\end{equation}
\end{subequations}
The last term of this variation results directly from varying $\sqrt{q}$. The remaining part can be written as a total covariant derivative of the third-rank tensor in Eq.~\eqref{eq:tensor-upsilon}. In contrast, it is impossible to write the variations of the terms $(\sqrt{q}/\kappa)a^cD_cu$ and $\sqrt{q}/(2\kappa)a^cD_cs^{\mathbf{nn}}$ in Eq.~\eqref{eq:bulk-hamiltonian-u} and Eq.~\eqref{eq:bulk-hamiltonian-snn}, respectively, in a similar form. This is probably the reason for why the $s^{ab}$ sector does not require an additional boundary term to match the field equations with the projection of the modified Einstein equations \cite{Bailey:2006fd} onto $\Sigma_t$.

The variation of the terms depending on Lie derivatives of the background tensor is
\begin{widetext}
\begin{align}
\label{eq:variations-momentum-terms-sab}
\delta\int_{\Sigma_t}\mathrm{d}^3y\,\left(P_{cd}-\frac{P}{2}q_{cd}\right)\mathcal{L}_ms^{cd}&=\int_{\Sigma_t}\mathrm{d}^3y\,\bigg\{P^a_{\phantom{a}c}\mathcal{L}_ms^{cb}+P^b_{\phantom{b}c}\mathcal{L}_ms^{ca}-\frac{1}{2}(P^{ab}q_{cd}\mathcal{L}_ms^{cd}+P\mathcal{L}_ms^{ab}) \notag \\
&\phantom{{}={}}\hspace{1.4cm}-\left(P_{cd}-\frac{P}{2}q_{cd}\right)\Big[N^{(a}\partial^{b)}s^{cd}-(\partial^{(a}N^c)s^{b)d}-(\partial^{(a}N^d)s^{b)c}\Big]\bigg\}\delta q_{ab}\,.
\end{align}
Varying the contributions involving the canonical momentum is lengthy, but \textit{xTensor} provides the result in a straightforward manner:
\begin{align}
\label{eq:variations-lie-derivative-terms-sab}
\delta\int_{\Sigma_t}\mathrm{d}^3y\,&\frac{2\kappa N}{\sqrt{q}}\left[P^{cd}P_{cd}-\frac{1}{2}(1-s^c_{\phantom{c}c})P^2-2s^{cd}(PP_{cd}-P_c^{\phantom{c}e}P_{de})\right] \notag \displaybreak[0]\\
&=\int_{\Sigma_t}\mathrm{d}^3y\,\frac{\kappa N}{2\sqrt{q}}\Big(\left[(1-s^c_{\phantom{c}c})P^2-2P^{cd}P_{cd}+4s^{cd}(PP_{cd}-P_c^{\phantom{c}e}P_{de})\right]q^{ab} \notag \displaybreak[0]\\
&\phantom{{}={}}\hspace{2.3cm}+8\Big[P^{ac}P_c^{\phantom{c}b}+s^{ac}P_{cd}P^{db}+s^{bc}P_{cd}P^{da}-s^{ac}PP_c^{\phantom{c}b}-s^{bc}PP_c^{\phantom{c}a} \notag \displaybreak[0]\\
&\phantom{{}={}}\hspace{2.3cm}+s^{cd}(P^a_{\phantom{a}c}P_d^{\phantom{d}b}-P_{cd}P^{ab})\Big]+2P^2s^{ab}-4(1-s^c_{\phantom{c}c})PP^{ab}\Big)\delta q_{ab}\,.
\end{align}
\end{widetext}
Finally, we should not forget to vary the boundary term in the action, Eq.~\eqref{eq:boundary-term-lagrangian-sab}:
\begin{subequations}
\begin{align}
\label{eq:variation-boundary-action-sab}
\delta B_{S,2}&=\oint_{S_t} \mathrm{d}^2\theta\,
\frac{N\sqrt{\sigma}}{2\kappa} \sigma^{ab} r^c  D_c \delta q_{ab}\notag \\
&\phantom{{}={}}+\oint_{S_t} \mathrm{d}^2\theta\,
\frac{N\sqrt{\sigma}}{2\kappa} \sigma^m_{\phantom{m}i} \sigma^n_{\phantom{n}j} s^{ij} r^k  D_k \delta q_{mn}\,,
\end{align}
where we have used
\begin{equation}
\label{expK_ij}
\delta k_{ij}=\sigma^m_{\phantom{m}i} \sigma^n_{\phantom{n}j} s^{ij} r^k  D_k \delta q_{mn}\,.
\end{equation}
\end{subequations}
As we found for the $u$ and $s^{\mathbf{nn}}$ sectors, Eq.~\eqref{eq:variation-boundary-action-sab} neatly cancels the sum in Eq.~\eqref{eq:variation-total-boundary-term-sab}: $\delta B_{S,2}-\delta B_{R,2}=0$.

Now we are ready to compile Eqs.~\eqref{eq:variation-ricci-with-boundary-term} (with $\delta B_R$ omitted), \eqref{eq:variation-ricci-with-boundary-term-sab} (with $\delta B_{sR}$ discarded), \eqref{eq:variation-acceleration-terms-sab}, \eqref{eq:variations-momentum-terms-sab}, and \eqref{eq:variations-lie-derivative-terms-sab} as well as Eq.~\eqref{eq:variation-shift-vector-term} adapted to the current sector. After inserting these variations into the second line of Eq.~\eqref{eq:Gaction2} and dropping the integral over $\Sigma_t$, we can cast the dynamical part of the modified Einstein equations into the following form:
\begin{widetext}
\begin{align}
\label{eq:einstein-equations-sab}
\dot{P}^{ab}&=\frac{\sqrt{q}}{2\kappa}\left(-NG^{ab}+D^aD^bN-q^{ab}D_cD^cN\right)
+\frac{N\sqrt{q}}{4\kappa}\Big(q^{ab}\left[R_{cd}-(D_c+a_c)(D_d+a_d)+a_cD_d\right]s^{cd} \notag \displaybreak[0]\\
&\phantom{{}={}}+(D_c+a_c)\Big[(D^a+a^a)s^{bc}+(D^b+a^b)s^{ac}-(D^c+a^c)s^{ab}\Big]\Big)-D_c\bigg(\frac{N\sqrt{q}}{4\kappa}\bigg[2a^{(a}s^{b)c}-s^{ab}a^c+q^{ab}s^{dc}a_d\bigg]\bigg) \notag \displaybreak[0]\\
&\phantom{{}={}}+\frac{1}{2}(P^{ab}q_{cd}\mathcal{L}_ms^{cd}+P\mathcal{L}_ms^{ab})-(P^a_{\phantom{a}c}\mathcal{L}_ms^{cb}+P^b_{\phantom{b}c}\mathcal{L}_ms^{ca})+\left(P_{cd}-\frac{P}{2}q_{cd}\right)\Big[N^{(a}\partial^{b)}s^{cd}-(\partial^{(a}N^c)s^{b)d} \notag \displaybreak[0]\\
&\phantom{{}={}}-(\partial^{(a}N^d)s^{b)c}\Big]-\frac{\kappa N}{2\sqrt{q}}\Big(\left[(1-s^c_{\phantom{c}c})P^2-2P^{cd}P_{cd}+4s^{cd}(PP_{cd}-P_c^{\phantom{c}e}P_{de})\right]q^{ab}+8\Big[P^{ac}P_c^{\phantom{c}b}+s^{ac}P_{cd}P^{db} \notag \displaybreak[0]\\
&\phantom{{}={}}+s^{bc}P_{cd}P^{da}-s^{ac}PP_c^{\phantom{c}b}-s^{bc}PP_c^{\phantom{c}a}+s^{cd}(P^a_{\phantom{a}c}P_d^{\phantom{d}b}-P_{cd}P^{ab})\Big]+2P^2s^{ab}-4(1-s^c_{\phantom{c}c})PP^{ab}\Big) \notag \displaybreak[0]\\
&\phantom{{}={}}-\sqrt{q}D_c\left(\frac{2N^{(a}P^{b)c}-P^{ab}N^c}{\sqrt{q}}\right)\,.
\end{align}
\end{widetext}
The latter is a generalization of $(\vec{\boldsymbol{q}}^{*}\mathbf{J}_2)^{ij}=0$ given in Eq.~(38a) of Ref.~\cite{Reyes:2022mvm} 
to a nonzero shift vector. Equations~\eqref{eq:einstein-equations-u}, \eqref{eq:einstein-equations-snn}, and \eqref{eq:einstein-equations-sab} 
completely govern the dynamics of the $u$, $s^{\mathbf{nn}}$, and $s^{ab}$ sectors of the modified-gravity theory based 
on Eq.~\eqref{eq:modified-EH}. The complexity of Eq.~\eqref{eq:einstein-equations-sab} illustrates the challenge of dealing 
with all sectors simultaneously, which is a manifestation of the profoundly nonlinear character of Eq.~\eqref{eq:modified-EH}. At the 
moment the best strategy seems to separate the sectors from each other in phenomenological studies. 
\section{Final Remarks}
\label{sec:FinalRemarks}
In this work, we have investigated a modification of GR governed by the $u$- and $s^{\mu\nu}$-type 
background fields contained in the minimal gravitational SME. The background fields were assumed to be 
nondynamical, which implies diffeomorphism breaking. Having carried out the ADM decomposition
of this theory in previous articles, our current focus was on a rigorous treatment of the 
gravitational boundary terms, which are unavoidable in this context.

To do so, we decomposed the spacetime boundary into two spacelike and one timelike hypersurface. As a 
consequence, the modified GHY boundary term split into three parts, each evaluated on one of the hypersurfaces 
previously referred to. Treating total-derivative terms in the action suitably canceled the 
extended GHY boundary term on the spacelike hypersurfaces. Foliating the timelike part of the boundary properly 
into two-dimensional hypersurfaces $S_t$, the remaining boundary contributions neatly combined 
to give rise to boundary terms on $S_t$. This procedure led us to the ADM-decomposed action of
Eq.~\eqref{eq:action-G-complete}, which is one of our central results.

Variations of the boundary term on $S_t$ for the intrinsic metric were
demonstrated to compensate further boundary terms originating from varying the Ricci 
scalar and Ricci tensor, respectively. Compiling the variations of each contribution in the
action for the intrinsic metric implied the dynamical field equations stated in Eqs.~\eqref{eq:einstein-equations-u},
\eqref{eq:einstein-equations-snn}, and \eqref{eq:einstein-equations-sab} for each
of the three sectors of the ADM-decomposed modified-gravity theory.

The only caveat was that for $u$ and $s^{\mathbf{nn}}$ further boundary terms depending explicitly
on the ADM acceleration had to be introduced to match the dynamical field equations with 
the modified Einstein equations on $\mathcal{M}$ of Ref.~\cite{Bailey:2006fd}, projected onto $\Sigma_t$.
A bonus of this new analysis is that it generalizes some of the findings in our previous 
paper \cite{Reyes:2022mvm} to a nonzero shift vector.

The formalism presented and results obtained are a well-suited starting point for phenomenology 
in black-hole physics affected by diffeomorphism violation. Moreover, from a theoretical viewpoint
they show that explicit diffeomorphism violation in gravity does not necessarily imply internal 
inconsistencies --- at least not at the level studied here and in our previous 
papers~\cite{Reyes:2021cpx,Reyes:2022mvm}. Time will show whether or not this 
conclusion can be upheld under different criteria.

\section{Acknowledgments}
It is a pleasure to thank P.~Sundell for valuable comments and for pointing out several important references
as well as Y.~Bonder, who informed us about another paper significant for your analysis.
C.M.R acknowledges partial support by the research project Fondecyt Regular 1191553 and would like to thank 
the post-graduate physics program at the Universidade Federal do Maranh\~{a}o (UFMA), S\~{a}o 
Lu\'{i}s, Brazil, where a part of this research was carried out, as well as M.M.~Ferreira, Jr.
and M.~Schreck for their kind hospitality. M.S. 
is indebted to FAPEMA Universal 00830/19, CNPq Produtividade 312201/2018-4, and CAPES/Finance Code 001.

%
%
%
%
%
%
%

\end{document}